\documentclass{jfm}

\usepackage{graphicx}
\usepackage{subfig}
\usepackage{newtxtext}
\usepackage{newtxmath}
\usepackage{natbib}
\usepackage{float}
\usepackage{makecell}

\usepackage{cite}
\usepackage{mathtools,amsmath,arraycols,array,textcomp}
\usepackage{booktabs,multirow,arydshln}
\usepackage{hyperref,cleveref}
\usepackage{float}

\usepackage{caption}
\usepackage{color} %
\usepackage{arydshln}

\hypersetup{
    colorlinks = true,
    urlcolor   = blue,
    citecolor  = black,
}

\newcommand{\RomanNumeralCaps}[1]
\linenumbers

\title{Dimensional homogeneity constrained gene expression programming for discovering governing equations}

\author{Wenjun Ma\aff{1},
  Jun Zhang\aff{1}
  \corresp{\email{jun.zhang@buaa.edu.cn}},
  Kaikai Feng\aff{1},
Haoyun Xing\aff{1}
\and  Dongsheng Wen \aff{1}}

\affiliation{\aff{1}School of Aeronautic Science and Engineering, Beihang University, Beijing, 100191, China}

\begin{document}
\maketitle

\begin{abstract}
  Data-driven discovery of governing equations is of great significance for helping us understand intrinsic mechanisms and build physical models. Recently, numerous highly innovative algorithms have emerged, aimed at inversely discovering the underlying governing equations from data, such as sparse regression-based methods and symbolic regression-based methods. Along this direction, a novel dimensional homogeneity constrained gene expression programming (DHC-GEP) method is proposed in this work. DHC-GEP simultaneously discovers the forms and coefficients of functions using basic mathematical operators and physical variables, without requiring pre-assumed candidate functions. The constraint of dimensional homogeneity is capable of filtering out the overfitting equations effectively. The key advantages of DHC-GEP compared to Original-GEP, including being more robust to hyperparameters, the noise level and the size of datasets, are demonstrated on two benchmark studies. Furthermore, DHC-GEP is employed to discover the unknown constitutive relations of two representative non-equilibrium flows. Galilean invariance and the second law of thermodynamics are imposed as constraints to enhance the reliability of the discovered constitutive relations. Comparisons, both quantitative and qualitative, indicate that the derived constitutive relations are more accurate than the conventional Burnett equations in a wide range of Knudsen number and Mach number, and are also applicable to the cases beyond the parameter space of the training data.
\end{abstract}



\section{Introduction}
\label{sec:intro}

Machine-learning-assisted modelling has become a new paradigm of research in a variety of scientific and engineering disciplines \citep{brunton2020machine,weinan2021dawning}. A very incomplete list includes \citet{ling2016reynolds}, \citet{koch2018mutual}, \citet{bergen2019machine}, \citet{sengupta2020ensembling}, \citet{ma2021non}, \citet{guastoni2021convolutional}, \citet{park2021toward}, \citet{karniadakis2021physics}, \citet{yu2022kinetic} and \citet{juniper2023machine}, the vast majority of which present improved performance, but a prominent critique is that the resulting models are “black boxes”. They cannot be explicitly expressed in mathematical forms. This not only sacrifices interpretability but also makes the resulting models difficult to disseminate between end users \citep{beetham2020formulating}.

In contrast to the aforementioned studies, \citet{bongard2007automated} and \citet{schmidt2009distilling} proposed using stratified symbolic regression (SR) and genetic programming (GP) to discover governing equations from data in low-dimensional systems. While this approach is innovative, it is difficult to scale up to high-dimensional systems. GP encodes equations using the nonlinear parse trees with varying sizes and shapes, which can become bloated in high-dimensional problems, making evolutions computationally expensive \citep{vaddireddy2020feature}. Subsequently, \citet{brunton2016discovering} proposed a seminal work called sparse identification of nonlinear dynamics (SINDy), which employs sparse regression to identify the most informative subset from a large predetermined library of candidate functions and determines the corresponding coefficients. SINDy is advantageous in deriving models that are explicit and concise, and has been widely used to discover equations in the form of first-order ODEs, alternatively with linear embedding \citep{lusch2018deep,champion2019data}, for applications in fluid systems \citep{loiseau2018constrained,loiseau2018sparse,zhang2020data}, predictive control of nonlinear dynamics \citep{kaiser2018sparse}, and multi-time-scale systems \citep{champion2019discovery}. To date, there have been many extensions to SINDy, such as partial differential equation functional identification of nonlinear dynamic (PDE-FIND) \citep{rudy2017data, schaeffer2017learning}, implicit sparse regression \citep{mangan2016inferring,kaheman2020sindy}, physics-constrained sparse regression \citep{loiseau2018constrained}, and sparse relaxed regularized regression \citep{zheng2018unified}. Basically, the sparse regression-based methods are confronted with two issues: the difficulty in accurately computing the derivative of noisy data and the requirement that all variables in the equation are observable. \citet{gurevich2019robust}, \citet{reinbold2020using} and \citet{alves2022data} employed the weak formulation of differential equations to decrease the noise sensitivity and eliminate the dependence on unobservable variables. Furthermore, \citet{reinbold2019data}, \citet{reinbold2021robust} and \citet{gurevich2021learning} considered three appropriate physical constraints, including locality, smoothness, and symmetries, to dramatically constrain the size of candidate library to be concise and effective.

More recently, in addition to sparse regression-based methods, two promising categories of data-driven methods have been proposed for discovering explicit models. The first category is the neural network-based method, such as PDE-Net \citep{long2018pde,long2019pde} and equation learner (${\mathrm{EQL}}$) \citep{sahoo2018learning}. PDE-Net approximates differential operators with convolutions and employs a symbolic multilayer neural network for model recovery, resulting in high expressivity and flexibility. EQL uses a special neural network structure whose activation functions are symbolic operators, and was demonstrated that the derived models can be generalized to the parameter spaces not covered by the training dataset. However, both neural network-based methods are criticized for the resulting equations being overly complex. 

Another category is gene expression programming (GEP)-based methods \citep{ferreira2001gene,vaddireddy2020feature,xing2022using}, which learn the forms of functions and their corresponding coefficients concurrently. The preselected elements for GEP include only mathematical operators, physical constants, and physical variables. The resulting equations are constructed by randomly combining basic elements while satisfying the syntactic requirements of the mathematical expression, rather than by linearly combining the predetermined candidate functions. Unlike the GP method used in \citet{schmidt2009distilling}, GEP encodes equations with fixed length linear strings that have unfixed open reading frames (ORFs). This feature of fixed length prevents bloating issues and excessive computational costs when dealing with complex problems. Additionally, variable ORFs ensure the variety of expression products, thus providing strong expressivity. Moreover, GEP performs a global exploration in the space of mathematical expressions, tending to obtain good results in a reasonable time. Therefore, in terms of data fitting, GEP has almost all the advantages of the aforementioned methods, featuring explicitness, concision, enhanced expressivity and flexibility without assuming the forms of candidate functions. Generally, GEP endeavors to discover the equations with less error; however, dimensional homogeneity cannot be guaranteed, particularly for problems with a variety of variables. Without any constraints or assumptions on the function forms, it is highly probably to obtain some overfitting and unphysical equations, which are sensitive to hyperparameters.

Considering the pros and cons of GEP, we propose a novel dimensional homogeneity constrained GEP (DHC-GEP) method for discovering governing equations. To the best of our knowledge, this is the first time that the constraint of dimensional homogeneity has been introduced to GEP. The constraint is implemented via a dimensional verification process before evaluating loss, without altering the fundamental features of Original-GEP \citep{ferreira2001gene}, including the structure of chromosomes, the rules of expression, selection, and genetic operators. Therefore, DHC-GEP inherits all the advantages of Original-GEP. More importantly, through two benchmark studies, we demonstrate that DHC-GEP has three critical improvements over Original-GEP: a) robust to the size and noise level of datasets, b) less sensitive to hyperparameters, and c) lower computational costs. 

\begin{figure} 
\centering  
\includegraphics[width=13.5cm]{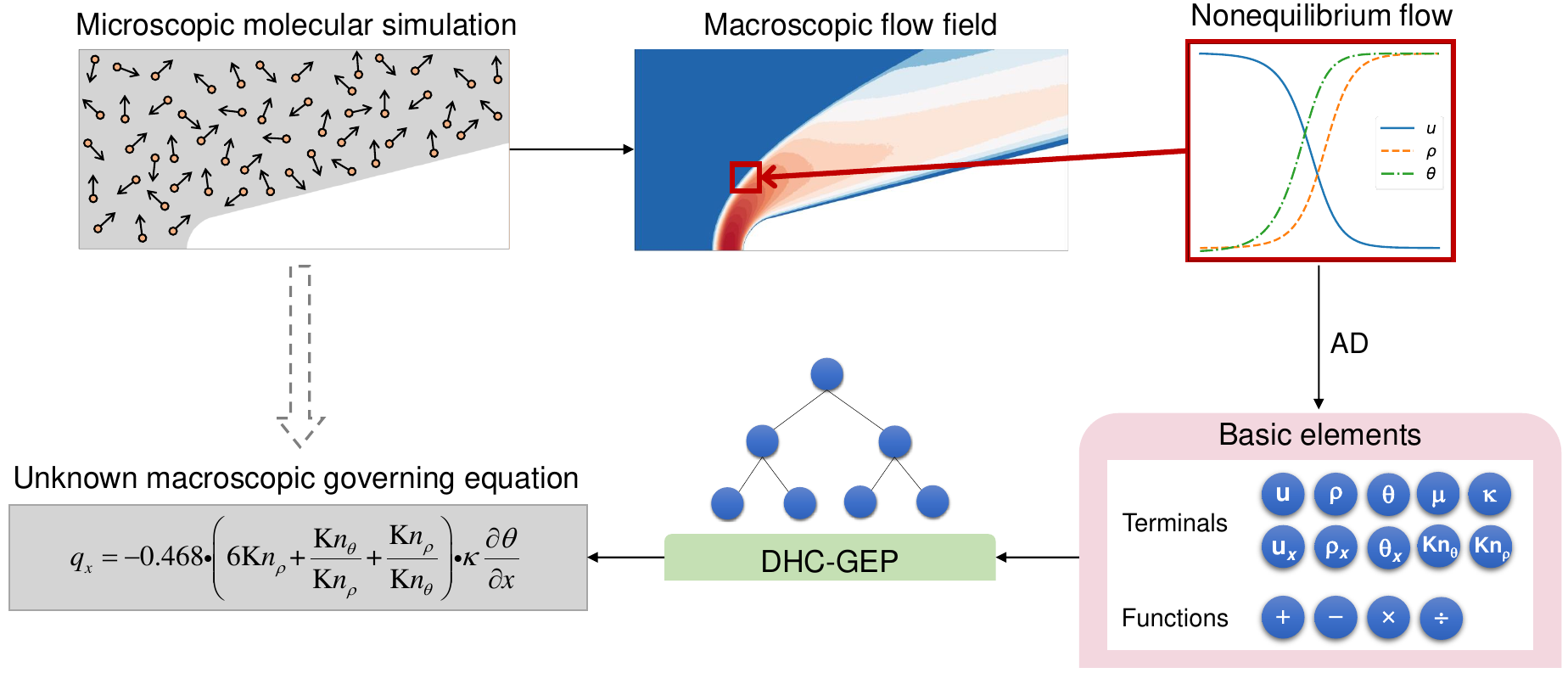}
\caption{Flowchart of discovering unknown governing equations from data. First, macroscopic flow field data is generated via microscopic molecular simulation. Then, the basic elements preselected for DHC-GEP are constituted with state variables, mathematical operators and other variables of concern. Taking the one-dimensional shock wave case as an example, the terminal set involves fundamental physical parameters (viscosity ($\mu $), heat conductivity ($\kappa $)), state variables (velocity ($u$), density ($\rho $), temperature ($\theta $)), gradient terms ($u_x$, $\rho_x$, $\theta_x$), and gradient-length local (GLL) Knudsen number (${Kn}_{\rho}$, ${Kn}_{\theta}$). The target variable is the heat flux in $x$ direction ($q_x$). The gradient terms are computed with auto differentiation (AD) based on neural networks, which is introduced in Appendix \ref{AD}. Finally, based on the terminal and function sets, DHC-GEP conducts a global search in the space of mathematical expressions until a satisfactory equation that fits well with the target variable is obtained.} 
\label{fig1}
\end{figure}

Furthermore, we extend the application of DHC-GEP to discover the unknown constitutive relations for non-equilibrium flows, including one-dimensional shock wave and rarefied Poiseuille flow. The conventional governing equations for fluid flows are the Navier-Stokes-Fourier (NSF) equations, which are derived based on the conservation of mass, momentum and energy, as well as the empirical assumptions of linear constitutive relations for the viscous stress and heat flux. Note that in strong non-equilibrium flows, these linear constitutive relations breakdown, and thus NSF equations are no longer applicable. Although high-order constitutive relations can be derived based on kinetic theory \citep{chapman1990mathematical}, such as Burnett equations \citep{burnett1936distribution}, their applicability is still very limited. Instead, our data-driven strategy is to derive the unknown constitutive relations from the data generated with molecular simulations. A general flowchart is shown in figure \ref{fig1}. Considering that constitutive relations describe the local transport mechanisms of momentum and energy, we regard local non-equilibrium parameters as key factors, and meticulously select the imported variables to satisfy the Galilean invariance \citep{han2019uniformly,huang2021learning}. Besides, the constraint of the second law of thermodynamics is embedded by adding an additional loss term, which is related to entropy production, to the loss function. The derived equations are more accurate than conventional Burnett equations over a wide range of Knudsen number and Mach number.  

The remainder of this paper is organized as follows. In §\,\ref{sec:DSMC}, we briefly introduce the molecular simulation method, i.e., the direct simulation Monte Carlo (DSMC) method. In §\,\ref{sec:DHC-GEP}, the DHC-GEP method is introduced in detail. In §\,\ref{sec:DOB}, we demonstrate the improved performance of DHC-GEP on two benchmark studies. Then, in §\,\ref{sec:AODUCR}, we extend its applications to discovering unknown constitutive relations for two non-equilibrium flows. Conclusions and discussions are drawn in §\,\ref{sec:CAD}.

\section{Methodology}
\label{sec:Methodology}

\subsection {Direct simulation Monte Carlo (DSMC)}
\label{sec:DSMC}

Generally, the Knudsen number ($Kn$), which is defined as the ratio of molecular mean free path to the characteristic length scale of system, is used to classify various flow regimes, with the continuum regime being in the range of $Kn \le 0.01$, the slip regime being $0.01 < Kn \le 0.1$, the transition regime being $0.1 < Kn \le 10$, and the free molecular flow being $Kn > 10$. While the computational fluid dynamics (CFD) methods based on the Navier-Stokes-Fourier (NSF) equations have been successfully applied to the continuum regime and can also be used for the slip regime with appropriate slip boundary conditions, they cannot accurately simulate transitional and free molecular gas flows. On the contrary, the Boltzmann equation can describe gas flows in all flow regimes, which can be written as 
\begin{equation}
    \frac{{\partial f}}{{\partial t}} + {\mathbf{c}}\frac{{\partial f}}{{\partial {\mathbf{x}}}} + {\mathbf{G}}\frac{{\partial f}}{{\partial {\mathbf{c}}}} = S,
\label{Boltzmann equation}
\end{equation}
Here, $f$ is the molecular velocity distribution function, which represents the density of molecules in phase-space (i.e., the space spanned by position and velocity). $\mathbf{c}$ is molecular velocity, $\mathbf{G}$ is external force, and $S$ is a complex integral term, which represents the change of distribution function due to the interaction between molecules. Owing to the presence of $S$, the Boltzmann equation is difficult to solve numerically. 

DSMC is a stochastic molecule-based method that approximates the molecular velocity distribution function in the Boltzmann equation with simulation molecules \citep{oran1998direct}. It has been demonstrated that DSMC converges to the solution of the Boltzmann equation in the limit of a large number of simulation molecules \citep{wagner1992convergence}. In DSMC, each simulation molecule represents $F$ real molecules, and $F$ is the so-called simulation ratio. DSMC tracks the simulation molecules as they move, collide with other molecules and reflect from boundaries. The macroscopic physical quantities are obtained via sampling corresponding molecular information and making an average at the sampling cells. Specifically, density ($\rho$) and macroscopic velocity ($v_i$) are computed with
\begin{equation}
    \rho =\frac{mFN_{p}}{V_{cell} } \mathrm{,} \;\mathrm{and}\;v_{i} =\frac{1}{N_p}\sum_{cell}c_i,
\label{DSMC_rho_u}
\end{equation}
where $m$ is molecular mass, $N_p$ is the number of simulation molecules in the sampling cell, and $V_{cell}$ is the volume of the sampling cell. The velocity of each molecule can be regarded as a sum of two parts, i.e., the macroscopic velocity of the cell to which it belongs and the molecular thermal velocity (which is defined as $C_i=c_i-v_i$). Based on the molecular thermal velocity, temperature ($T$) is computed with
\begin{equation}
    T=\frac{1}{3k_{B}N_{p}}\sum_{cell}mC^{2},
\label{DSMC_T}
\end{equation}
where $k_B$ is the Boltzmann constant, and $C^2=C_1^2+C_2^2+C_3^2$. Pressure tensor ($p_{ij}$) is computed with
\begin{equation}
    p_{ij}=\frac{mF}{V_{cell}}\left ( {\sum_{cell}c_ic_j}-N_pv_iv_j \right ),
\label{DSMC_p_ij}
\end{equation}
and is generally split into its trace and trace-free parts, i.e., pressure ($p$) and stress tensor ($\tau_{ij}$), $p_{ij}=p\delta_{ij}+\tau_{ij}$. Here, $\delta_{ij}$ is Kronecker delta \citep{heinbockel2001introduction}. Therefore, pressure ($p$) and stress tensor ($\tau_{ij}$) are computed with
\begin{equation}
    p=\frac{1}{3} \left ( p_{11}+p_{22}+p_{33} \right ) \mathrm{,}\; \mathrm{and } \;\tau_{ij}=p_{ij}-p\delta_{ij}.
\label{DSMC_p_tau_ij}
\end{equation}
The heat flux ($q_i$) is computed with
\begin{equation}
    q_i=\frac{F}{2V_{cell}}\sum_{cell}mC^{2}C_i .
\label{DSMC_q_i}
\end{equation}
The local viscosity coefficient is computed with 
\begin{equation}
    \mu =\mu _0\left ( \frac{T }{T _0}  \right ) ^\omega  ,
\label{viscosity_eq}
\end{equation}
where $\mu_0 = 2.117\times 10^{-5} \mathrm{kg\cdot m^{-1}\cdot s^{-1}}  $ is the reference viscosity at the reference temperature $T_0 = 273 \mathrm{K} $. $\omega$ is the viscosity exponent and is determined by the molecular model employed. For hard sphere (HS) model \citep{bird1994molecular}, Maxwell molecule model \citep{bird1994molecular}, and variable hard sphere (VHS) model \citep{bird1981monte}, $\omega$ is equal to 0.5, 1.0, and 0.81, respectively. The local heat conductivity coefficient is computed according to the local viscosity coefficient using
\begin{equation}
    \kappa =\frac{5}{2}\frac{\mu }{\mathrm{Pr} },
\label{heat_conductivity_eq}
\end{equation}
where $\mathrm{Pr}$ is the Prandtl number and is equal to $2/3$ for monatomic gas.

For a specific application, DSMC first initializes the simulation molecules according to the initial distributions of density, temperature, and macroscopic velocity. Then, the molecular motions and inter-molecular collisions are sequentially conducted in each computational time step. The molecular motions are implemented in a deterministic way. Each molecule moves ballistically from its original position to a new position, and the displacement is equal to the product of its velocity times the time step. If the trajectory crosses any boundaries, an appropriate gas-wall interaction model would be employed to determine the reflected velocity. Common gas-wall interaction models include specular, diffuse and Maxwell reflection models. In our simulations, diffuse model is employed. The inter-molecular collisions are implemented in a probabilistic way. Among the several algorithms for modelling inter-molecular collisions, the no-time-counter (NTC) technique \citep{bird1994molecular} is the most widely used. In NTC, any two molecules in the same computational cell are selected as a collision pair with a probability that is proportional to the relative speed between them. Then, the post-collision velocities of molecules are determined depending on the molecular model employed. In this work we employ the HS model for the first two benchmark cases (diffusion flow and Taylor-Green vortex), the Maxwell molecule model for the shock wave case, and the VHS model for the Poiseuille flow case. It is noteworthy that because the molecular motions and inter-molecular collisions are conducted in a decoupled manner, the time step in DSMC needs to be smaller than the molecular mean collision time, and the cell size for the selection of collision pairs needs to be smaller than the molecular mean free path.

DSMC has been successfully applied to simulate gas flows in the whole flow regimes, and the results have been well validated \citep{sun2002direct,stefanov2002rayleigh,zhang2010effects,gallis2017molecular}. The advantage of the DSMC method is that there is no need to assume any macroscopic governing equations a priori. The macroscopic quantities, such as density and velocity, are obtained by sampling molecular information and making an average at the sampling cells. In this work, the aim of using data generated by DSMC is to prove the ability of DHC-GEP to discover the governing equations that could be unknown. 

\subsection{Dimensional homogeneity constrained gene expression programming (DHC-GEP)}
\label{sec:DHC-GEP}

\subsubsection{General framework}
\label{sec:General flowchart}

\begin{figure} 
\centering  
\subfloat{\includegraphics[width=13.5cm]{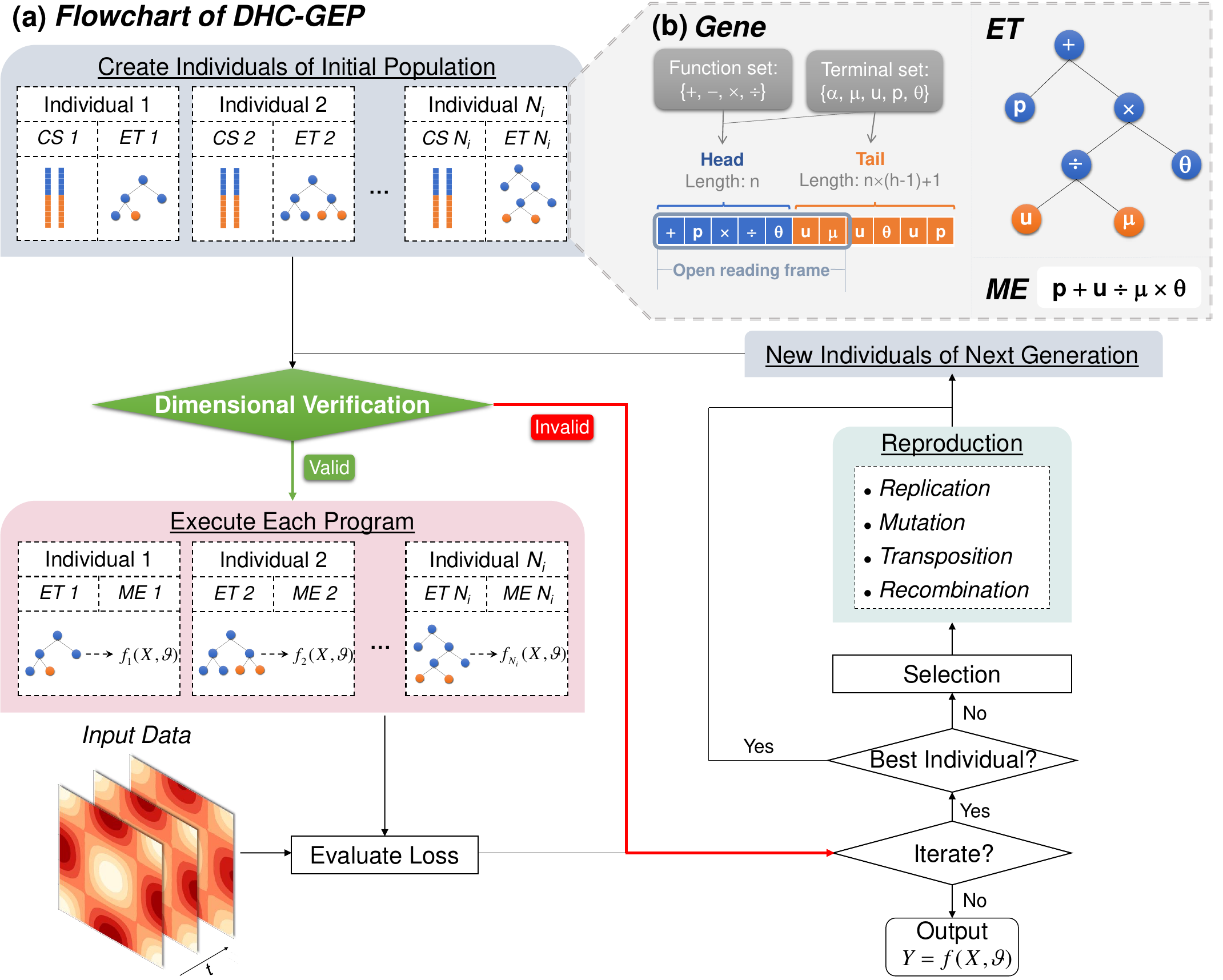}}

\subfloat{\includegraphics[width=13.5cm]{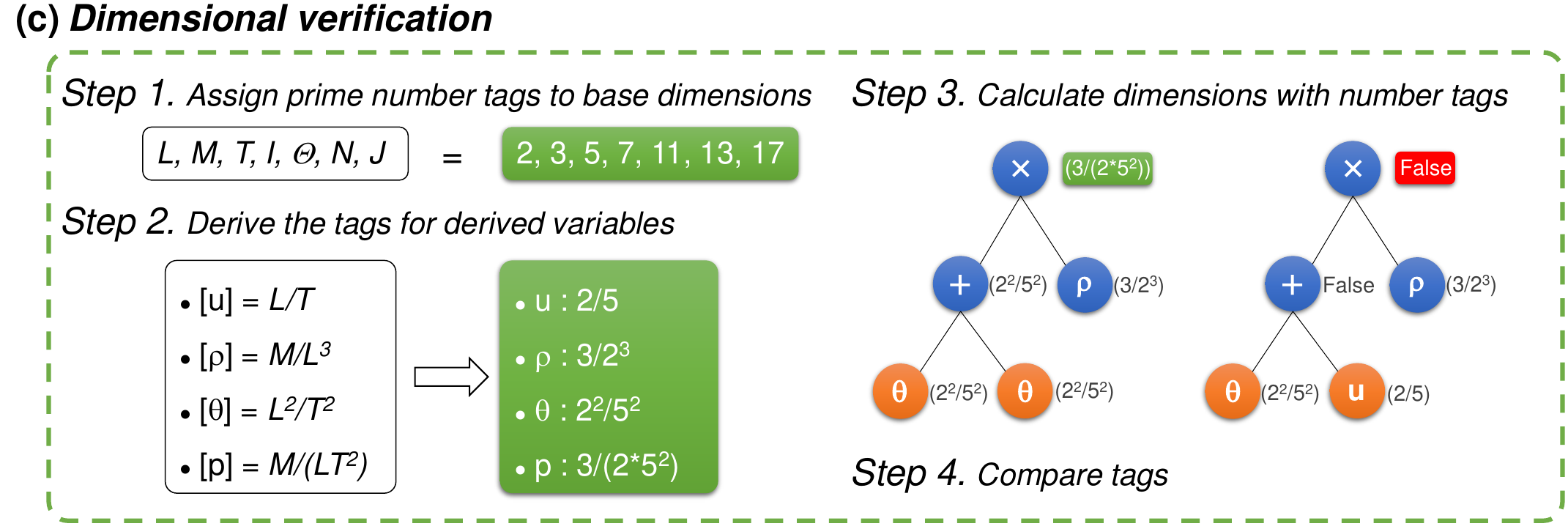}}
\caption{Main characteristics of DHC-GEP. (a) Flowchart of DHC-GEP. (b) Schematic diagram of a gene and its corresponding expression tree and mathematical expression. (c) Strategy of dimensional verification.} 
\label{fig2}
\end{figure}

\begin{figure}  
\centering  
\includegraphics[width=12cm]{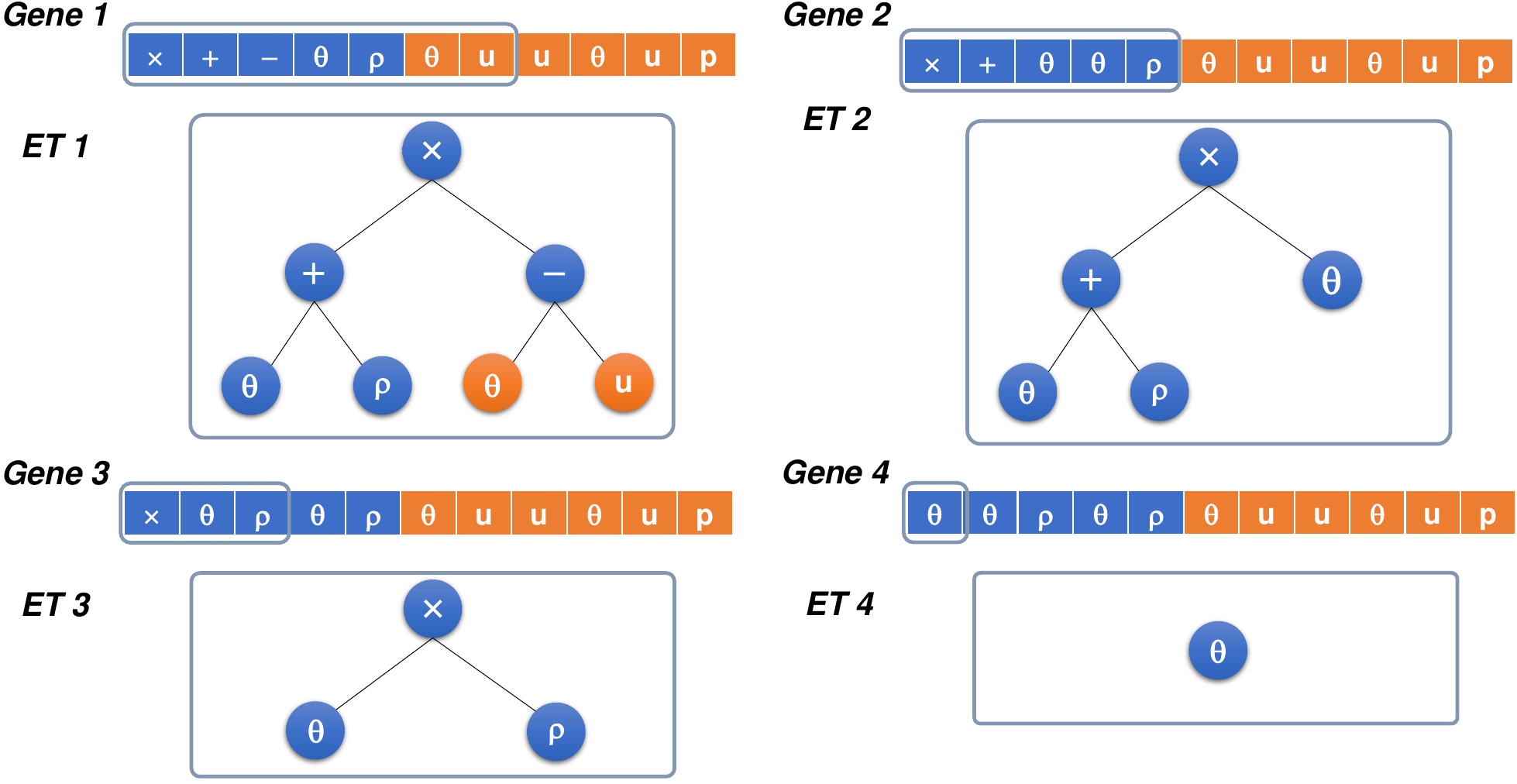}
\caption{Four genes with different ORFs.} 
\label{su_fig1}
\end{figure}

DHC-GEP is an enhanced gene expression programming method, which compiles mathematical expressions into chromosomes and iteratively discovers the equation that is suitable for describing the training data by mimicking the natural law of evolution. The flowchart of DHC-GEP is shown in figure \ref{fig2}(a). It starts with creating $N_i$ random individuals of initial population. Each individual has two forms: genotype (chromosome (CS)) and phenotype (expression tree (ET)). Phenotype is the expression product of genotype. Each chromosome is composed of one or more genes. A specific schematic diagram of gene is shown in figure \ref{fig2}(b). There are two parts in a gene, i.e., head and tail. The head consists of the symbols from the function set or terminal set, while the tail consists of only symbols from the terminal set. The function set and terminal set are both predefined according to the specific problem. For the problems in this work, the function set includes basic mathematical operators ($ + ,\; - ,\; \times ,\; \div $). If necessary for other problems, the function set can also include nonlinear functions such as sin, cos, and even user-defined functions. The terminal set includes the symbols of variables and physical constants. Taking the first symbol of gene as the root node, we can translate the gene into the expression tree (shown in figure \ref{fig2}(b)) through level-order traversal according to the argument requirement of each function. Note that the final four symbols are not expressed, and the region before them is called open reading frame (ORF). The length of ORF is variable, which is determined by the argument requirement of the function nodes in the head. On the contrary, the length of gene is fixed, preventing the individuals from bloating. Assuming the length of head is $n$, then the length of tail must be $n \times \left( {h - 1} \right) + 1$, where $h$ is the maximum number of arguments in the functions. Four genes with different ORFs are shown in figure \ref{su_fig1}. Although the four genes have the same length, the complexities of their expression products are different. Therefore, ORF is the decisive factor that ensures the diversity of expression products and the strong expressivity of DHC-GEP. This is also the reason that the length of head determines only the upper limit of complexity, not the lower limit. When discovering unknown governing equations, a relatively long head would be a good choice. Because with ORF, a long gene can represent either a simple equation or a complex equation.

\begin{figure}  
\centering  
\includegraphics[width=12cm]{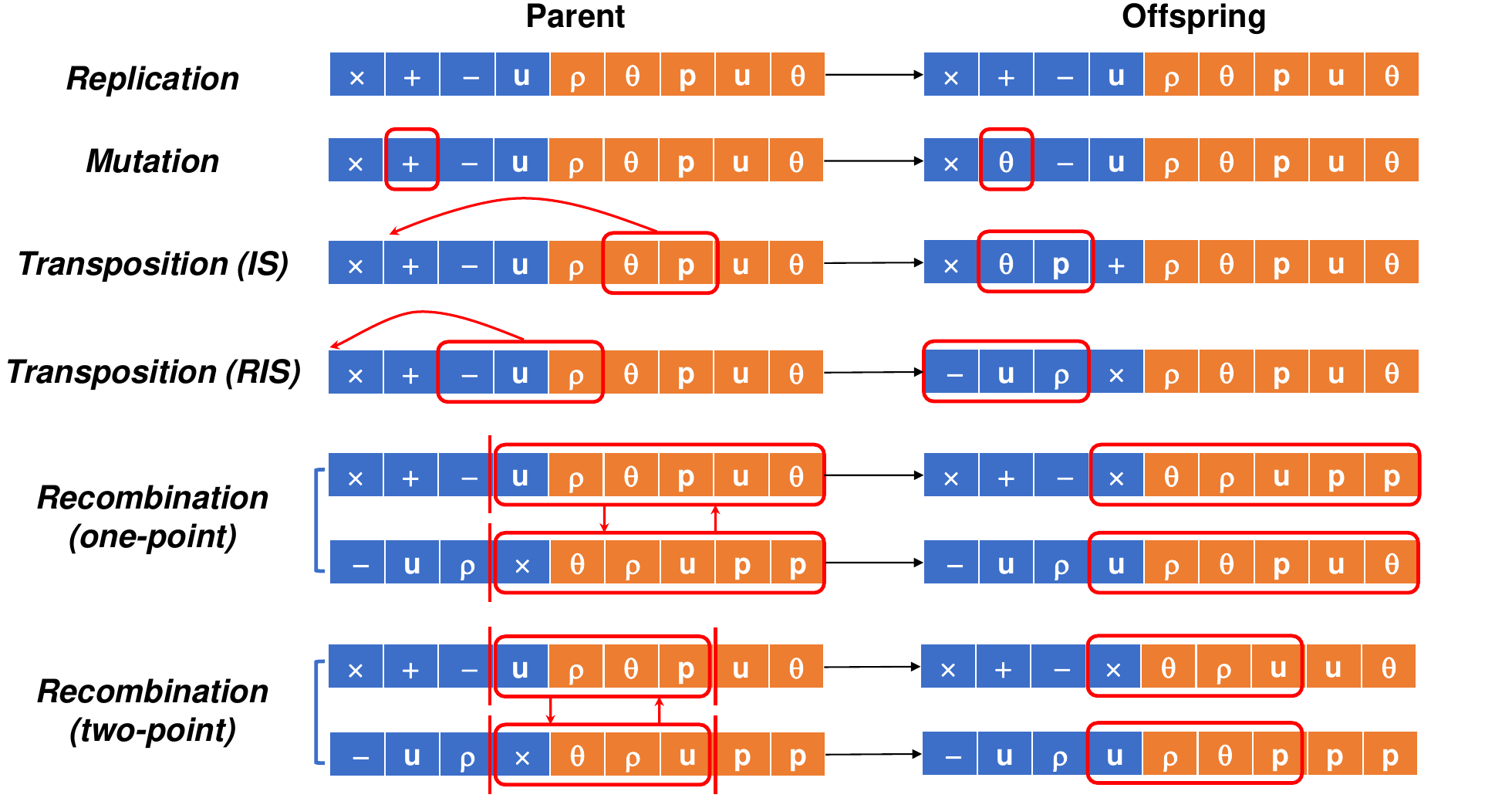}
\caption{Schematic diagrams of genetic operators.} 
\label{su_fig3}
\end{figure}

Then, via dimensional verification (which will be introduced in §\,\ref{sec:Dimensional verification}), all the individuals are classified into valid individuals and invalid individuals according to whether they satisfy the dimensional homogeneity. The valid individuals would be translated into mathematical expressions (MEs) and evaluated for losses with input data according to the loss function, which is defined using the mean relative error (MRE) as
\begin{equation}
    {\mathrm{L_{MRE}}} = \frac{1}{N}\sum\limits_{i = 1}^N {\left| {\frac{{{{\hat Y}_i} - {Y_i}}}{{{Y_i}}}} \right|} .
\label{eq16}
\end{equation}
Here, the variable with a superscript $\wedge$ is the predicted variable, and $N$ is the total number of data points. The invalid individuals would be directly assigned a significant loss and eliminated in the subsequent iterations. The best individual (with the lowest loss) is replicated to the next generation straightly. Other individuals are selected as superior individuals in the probabilities that are inversely proportional to their losses. Based on the superior individuals, the population of next generation is reproduced via genetic operators. Common genetic operators include replication, mutation, transposition and recombination. The schematic diagrams of them are shown in figure \ref{su_fig3}; the details are introduced as follows:
\begin{itemize}
    \item \textbf{Replication} is responsible for copying the selected superior individuals to the next generation. After replication, the population maintains the same size as the initial population.
    \item \textbf{Mutation} is a change in some symbols of genes. To maintain the structural organization of genes, the symbols in head can be changed into either functions or terminals, while the symbols in tail can only be changed into terminals. In addition, note that mutations in genes do not necessarily lead to changes in expression products, unless the mutation occurs in ORF.
    \item \textbf{Transposition} is the replacement of a gene segment with another position. Specifically, it includes insertion sequence (IS) transposition, root insertion sequence (RIS) transposition, and gene transposition. For IS transposition, a short segment with a function or terminal in the first position is transposed to the head of genes. The length of the transposed segment and the insertion site are randomly selected. The sequence before the insertion site remains unchanged. The sequence after the insertion site is shifted backwards as a whole, and the last symbols (as many as the length of the transposed segment) of the head are deleted. RIS transposition is similar to IS transposition, except that the first position of the transposed segment must be function, and the insertion site must be the root. Gene transposition occurs in the chromosomes that consist of multiple genes. During gene transposition, one gene switches position with another gene. 
    \item \textbf{Recombination} is an exchange of gene segments between two selected parent chromosomes, including one-point recombination, two-point recombination and gene recombination. For one-point recombination, one bond is randomly selected as the crossover point. Then, both selected chromosomes are cut at this point, and exchange the segments after the crossover point. For two-point recombination, two crossover points are selected, and the exchanged segments are between these two points. For gene recombination, the exchanged segment is an entire gene.
\end{itemize}

Note that the genetic operators, except replication, are not necessarily invoked in every chromosome, but are invoked with certain probabilities. In this work, the probabilities of the genetic operators being invoked refer to \citet{ferreira2006gene}, listed in table \ref{su_tab4}, which were concluded from various examples via the trial and error approach.

\begin{table}
\centering
\begin{tabular}{lc} 
Genetic operator          & Probabilities  \\ 
Mutation                  & 0.05           \\
Inversion                 & 0.1            \\
IS transposition          & 0.1            \\
RIS transposition         & 0.1            \\
Gene transposition        & 0.1            \\
One-point recombination   & 0.3            \\
Two-point recombination   & 0.2            \\
Gene recombination        & 0.1            \\
\end{tabular}
\caption{\label{su_tab4} Probabilities for genetic operators being invoked.}
\end{table}

The above processes are iteratively conducted until a satisfactory individual is obtained. Specifically, in benchmark studies, we manually terminate the evolution when the target equation is discovered. In cases of discovering the unknown constitutive relations, the evolution is terminated automatically if the optimal equation remains unchanged for 1,000 consecutive generations.

Other advanced knowledge is provided in Appendix, including the Dc domain (Appendix \ref{Dc domain}) for generating numerical constants and the linear scaling (Appendix \ref{linear scaling}) for discovering numerical coefficients.

\subsubsection{Dimensional verification}
\label{sec:Dimensional verification}

The dimensional verification is the additional process to implement the constraint of dimensional homogeneity, and is introduced as follows (a simplified version is shown in figure \ref{fig2}(c)). The general form of governing equation is assumed to be:
\begin{equation}
Y = f(X,\vartheta ).
\label{eq1}
\end{equation}
Here, $Y$ is the target variable, and $f(\cdot )$ is the mathematical expression coded by chromosome. $X$ and $\vartheta$  represent variables and physical constants, respectively. Dimensional homogeneity means that $f(\cdot )$ should have the same dimension as $Y$, and every part in $f(\cdot )$ should satisfy the constraint that the parameters for operators $ + $ or $ - $ must have the same dimensions.

Generally, dimensional verification includes calculating the dimension of $f(\cdot )$ and comparing it with the dimension of $Y$. For calculating dimensions, there are five principles that should be noted.
\begin{itemize}
    \item Computers cannot directly deal with symbolic operations, but only numerical operations.
    \item In the International System of Units (SI), there are seven base dimensions: length ($L$), mass ($M$), time ($T$), electric current ($I$), thermodynamic temperature ($\Theta $), amount of substance ($N$), and luminous intensity ($J$). The dimensions of any other physical quantities can be derived by powers, products, or quotients of these base dimensions.
    \item Base dimensions are independent of each other. Any dimension cannot be derived by other base dimensions.
    \item Physical quantities with different dimensions cannot be added or subtracted. Adding or subtracting the physical quantities with the same dimension does not change the dimension.
    \item When physical quantities are multiplied or divided, the corresponding dimensions are multiplied or divided equally.
\end{itemize}

Considering the above principles, we propose to assign number tags to physical quantities and replace the dimensional calculations with numerical calculations. For the seven base dimensions, the tags are 2, 3, 5, 7, 11, 13, and 17, respectively. These numbers are prime numbers, ensuring that the base dimensions are independent of each other. The tags for other physical quantities are derived according to their dimensions. For example, as the dimension of velocity is length divided by time ($L/T$), its tag is defined as $2/5$. It is noteworthy that the tags are always in the form of fractions rather than floats to avoid truncation errors. The tags for numerical constants and dimensionless quantities are set to $1$. In addition, according to the last two principles, we modify the operation rules of mathematical operators. The pseudo-codes for addition ($ + $) and multiplication ($ \times $) are provided in Algorithm 1 and Algorithm 2, respectively. Subtraction ($ - $) and division ( $ \div $) are defined in similar ways. With the number tags and modified mathematical operators, we can compute the dimension of each node in the expression tree from the bottom up. If the parameters for one function node ($ + $ or $ - $) does not have the same dimension, the individual is directly identified as invalid one. Otherwise, we can calculate the specific dimension of the individual, i.e., the number tag for the root node. Comparing it with the tag for the target variable, we can conclude whether the individual is dimensionally homogeneous.

\begin{table}
\centering
\begin{tabular}{cll} 
\multicolumn{3}{l}{\textbf{Algorithm 1} Modified addition function}                                                                                                                                               \\ 
\hline
1 & \multicolumn{2}{l}{\textbf{Input:} tags for the input variables, denoted by \textbf{a} and \textbf{b}}                                                                                               \\
2          & \multicolumn{2}{l}{Modified\_addition(\textbf{a},\textbf{b}):}                                                                                                                                       \\
3          &  & If \textbf{a} or \textbf{b} is bool value:\textcolor[rgb]{0.753,0.753,0.753}{~ ~ ~\textit{\# Indicates that this equation has been proved to be invalid at previous nodes.}}  \\
4          &  & ~ ~ Return False~ ~ ~\textit{\textcolor[rgb]{0.753,0.753,0.753}{\# Continue transfer the message that this equation is invalid.}}                                   \\
5          &  & If \textbf{a} is integer:~ ~ ~\textit{\textcolor[rgb]{0.753,0.753,0.753}{\# Indicates that the input variables is a numerical constant.}}                                                                        \\
6          &  & ~ ~ Assign 1/1 (i.e. 1 in the form of fraction) to \textbf{a} ~ ~ ~\textit{\textcolor[rgb]{0.753,0.753,0.753}{\# Assign the number tag 1 to the numerical constant.}}                                        \\
7          &  & If \textbf{b} is integer:~ ~ ~\textit{\textcolor[rgb]{0.753,0.753,0.753}{\# Indicates that the input variables is a numerical constant.}}                                                                        \\
8          &  & ~ ~ Assign 1/1 (i.e. 1 in the form of fraction) to \textbf{b}~ ~ ~\textit{\textcolor[rgb]{0.753,0.753,0.753}{\# Assign the number tag 1 to the numerical constant.}}                                        \\
9          &  & If \textbf{a} is equal to \textbf{b}:\textcolor[rgb]{0.753,0.753,0.753}{~ ~ ~\textit{\# Indicates that the input variables have the same dimensions.}}                                                 \\
10         &  & ~ ~ Return \textbf{a}~ ~ ~\textit{\textcolor[rgb]{0.753,0.753,0.753}{\# Return the tag for any input variables as the tag for this node.}}                                          \\
11         &  & Else:~ ~ ~\textit{\textcolor[rgb]{0.753,0.753,0.753}{\# Indicates that the input variables do not have the same dimensions.}}                                                                           \\
12         &  & ~ ~ Return False\textcolor[rgb]{0.753,0.753,0.753}{~ ~ ~\textit{\# Judge that this equation is invalid.}}                                                           \\
\hline
\end{tabular}
\end{table}

\begin{table}
\centering
\begin{tabular}{lll} 
\multicolumn{3}{l}{\textbf{Algorithm 2} Modified multiplication function}                                                                                                                                \\ 
\hline
1 & \multicolumn{2}{l}{\textbf{Input:} tags for the input variables, denoted by \textbf{a} and \textbf{b}}                                                                                                                 \\
2 & \multicolumn{2}{l}{Modified\_multiplication(\textbf{a},\textbf{b}):}                                                                                                                                 \\
3 &  & If \textbf{a} or \textbf{b} is bool value:~ ~ \textcolor[rgb]{0.753,0.753,0.753}{~\textit{\# Indicates that this equation has been proved to be invalid at previous nodes.}}  \\
4 &  & ~ ~ Return False~ ~ \textcolor[rgb]{0.753,0.753,0.753}{~\textit{\# Continue transfer the message that this equation is invalid.}}                                   \\
5 &  & If \textbf{a} is integer:~ ~ ~\textit{\textcolor[rgb]{0.753,0.753,0.753}{\# Indicates that the input variables is a numerical constant.}}                                                                        \\
6 &  & ~ ~ Assign 1/1 (i.e. 1 in the form of fraction) to \textbf{a}~ ~ \textcolor[rgb]{0.753,0.753,0.753}{~\textit{\# Assign the number tag 1 to the numerical constant.}}                                        \\
7 &  & If \textbf{b} is integer:~ ~ ~\textit{\textcolor[rgb]{0.753,0.753,0.753}{\# Indicates that the input variables is a numerical constant.}}                                                                        \\
8 &  & ~ ~ Assign 1/1 (i.e. 1 in the form of fraction) to \textbf{b}~ ~ ~\textit{\textcolor[rgb]{0.753,0.753,0.753}{\# Assign the number tag 1 to the numerical constant.}}                                        \\
9 &  & Return \textbf{a$\times$b }in the form of fraction~ ~ ~\textcolor[rgb]{0.753,0.753,0.753}{\makecell[l]{\# Return the product of the tags for the input variables as the \\ ~ tag for this node.}}                        \\
\end{tabular}
\label{AL2}
\end{table}

Note that the datasets and source codes used in this work are available on GitHub at \href{https://github.com/Wenjun-Ma/DHC-GEP}{https://github.com/Wenjun-Ma/DHC-GEP}.

\section{Demonstration on benchmark studies}
\label{sec:DOB}

\subsection {Diffusion equation}
\label{sec:diffusion flow}

\begin{figure}  
\centering  
\includegraphics[width=12cm]{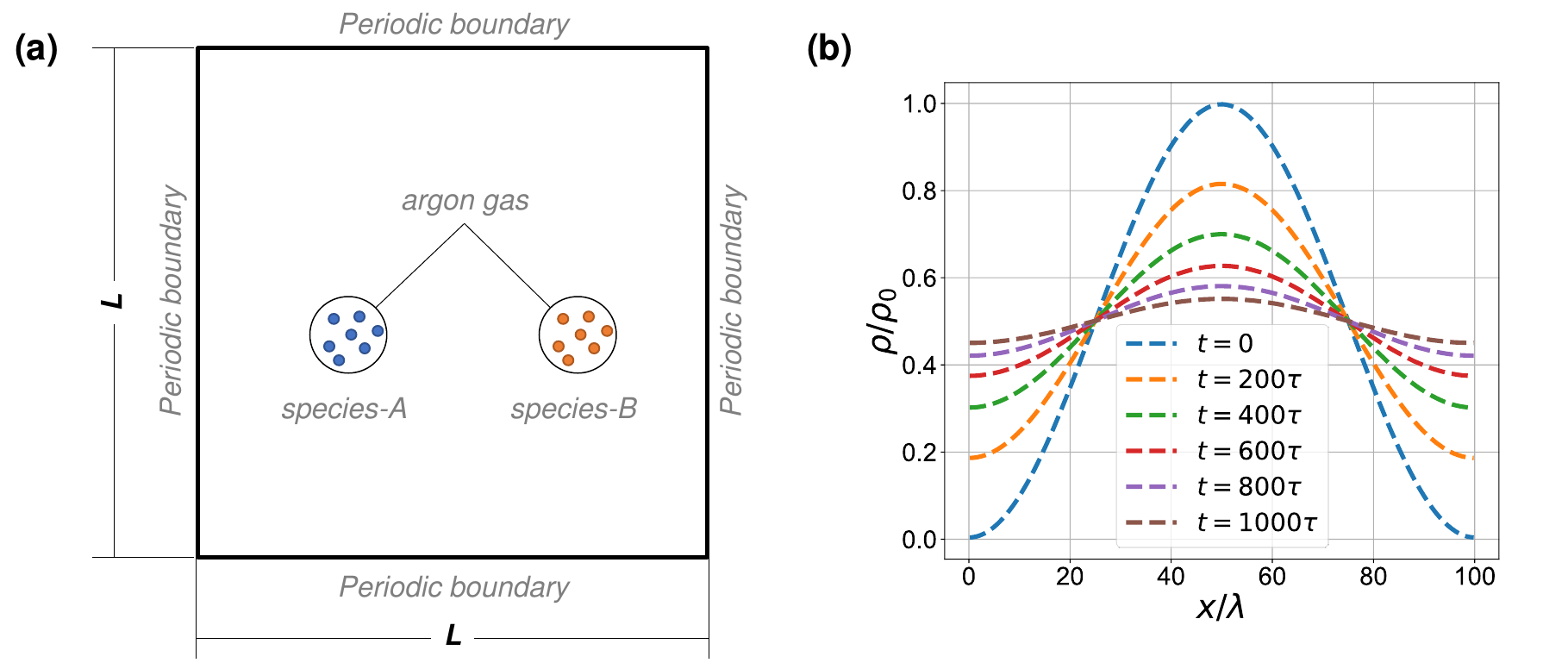}
\caption{(a) Schematic diagram of the computational domain for the diffusion flow. (b) Distributions of density at the instants of $t= 0,\;200\tau ,\;400\tau ,\;600\tau ,\;800\tau ,\;1000\tau $.} 
\label{su_fig5}
\end{figure}

We employ the DSMC method to simulate a two-dimensional diffusion flow in a square domain, as shown in figure \ref{su_fig5}(a). The side length of the square is $L = 100\lambda $, and $\lambda $ is the mean free path of argon gas molecules at standard condition (pressure $p = 1.01 \times {10^5}\;{\rm{Pa}}$ and temperature $T = 273\;{\rm K}$), i.e., $\lambda=6.25\times 10^{-8}\;\mathrm{m}$. As a result, the Knudsen number ($Kn = \lambda / L$) is 0.01. All the four sides are set to periodic boundaries. The simulation includes only argon gas, which is divided into two species, denoted as species-A and species-B, respectively. At the initial instant, the density distributions of these two species are ${\rho _A}{\rm{ = }}0.5{\rho _0}\left( {1 - \cos \left( {2\upi y/L} \right)} \right)$ and ${\rho _B}{\rm{ = }}0.5{\rho _0}\left( {1 + \cos \left( {2\upi y/L} \right)} \right)$, respectively. Here, ${\rho _0}=1.78\;{\rm kg} \cdot {{\rm m}^{ - 3}}$ is the density of argon gas at standard conditions, and $y$ is the spatial coordinate in the vertical direction. The initial macroscopic velocity is uniformly zero. During simulation, the macroscopic velocity and the total density of the two species remain uniform and unchanged, while the respective densities of these two species vary with space and time.

The computational domain is divided into $256 \times 256$ cells, which means that the cell size ($\Delta x,\Delta y$) is approximately $0.4\lambda $. In each cell, 4 000 simulation molecules are randomly distributed at the initial instant, with one simulation molecule representing $4\times 10^{6}$ real molecules. The initial thermal velocity for each simulation molecule is sampled randomly from a Maxwell distribution at 273 K. The computational time step ($\Delta t$) is $0.1\tau $, where $\tau $ is the mean collision time of argon gas molecules at standard condition with the definition of $\tau =\lambda /\bar{c}$. Here, $\bar{c}$ is the molecular mean velocity, i.e., $\bar{c} =\sqrt{8k_{B}T/\pi m} $, where $k_B$ and $m$ are the Boltzmann constant and molecular mass, respectively. 

Note that the density varies only along the vertical direction. We average the data along the horizontal direction to reduce the statistical errors. During simulation, we sample the macroscopic densities of species-A at the instants of $t = 0,\;10\tau ,\;20\tau ,...,\;1000\tau $. The temporal evolution of density distributions is shown in figure \ref{su_fig5}(b). The final dataset is composed of 25 856 ($256 \times 101$) data points that are distributed in 256 cells along the vertical direction at 101 instants. 

The theoretical governing equation of this flow is the diffusion equation, i.e., 
\begin{equation}
\frac{{\partial \rho }}{{\partial t}} = D\frac{{{\partial ^2}\rho }}{{\partial {y^2}}}.
\label{eq2}
\end{equation}
Here, $D$ is the diffusion coefficient. According to the Chapman-Enskog theory \citep{chapman1990mathematical}, the diffusion coefficient for an HS gas at the first-order approximation is
\begin{equation}
    D=\frac{3\pi }{16} \frac{\lambda ^{2} }{\tau } .
\label{diffusion coefficient}
\end{equation}
Therefore, for the argon gas in our simulation, $D$ equals $1.40 \times {10^{ - 5}}\;{{\rm m}^2}{{\rm s}^{ - 1}}$.

Before using the dataset generated by DSMC, we test DHC-GEP and Original-GEP on a clean dataset. The only difference between the DSMC dataset and the clean dataset lies in the target variable ($\frac{{\partial \rho }}{{\partial t}}$). In the clean dataset, $\frac{{\partial \rho }}{{\partial t}}$ is directly computed based on $D$ and $\frac{{{\partial ^2}\rho }}{{\partial {y^2}}}$, according to (\ref{eq2}). We define the function set as $\left\{ { + , - , \times , \div } \right\}$, and the terminal set as $\left\{ {\mu ,D,\frac{{\partial \rho }}{{\partial y}},\frac{{{\partial ^2}\rho }}{{\partial {y^2}}}{\rm{,}}\frac{{{\partial ^3}\rho }}{{\partial {y^3}}}} \right\}$. Here, $\mu$ is the viscosity coefficient, and is introduced as a distraction term. According to the Chapman-Enskog theory \citep{chapman1990mathematical}, $\mu$ equals $\frac{5\pi }{32} \frac{\lambda ^{2} }{\tau }\rho = 2.08  \times {10^{ - 5}}\;{\rm kg}\cdot {{\rm m}^{-1}}{{\rm s}^{ - 1}}$. All the hyperparameters are set to be the same for both GEP methods across all cases (see in Appendix \ref{appB}). The derived equations are shown in table \ref{tab2}. It can be observed that the derived equation of DHC-GEP is consistent with the theoretical equation, while the derived equation of Original-GEP is partially correct. Specifically, Original-GEP identifies the correct function form and coefficient ($1.40 \times {10^{ - 5}}$). In the view of data fitting, this is a correct result. However, Original-GEP fails to recognize that the coefficient ($1.40 \times {10^{ - 5}}$) preceding the diffusion term represents the diffusion coefficient. Consequently, while the resulting equation may be numerically accurate, it lacks complete physical significance. For another diffusion flow with a different diffusion coefficient, the derived equation of Original-GEP is no longer valid, while the derived equation of DHC-GEP is generally applicable.

\begin{table}
\centering
\renewcommand\arraystretch{2.4}
\resizebox{\textwidth}{3.0cm}{
\begin{tabular}{llllcc} 
Case                                    & Theoretical equation & Data                           & Method       & Derived equation & Loss   \\ 
\hdashline[1pt/1pt]
\multirow{4}{*}{Diffusion problem}      & \multirow{4}{*}{$\begin{array}{l}
    \frac{{\partial \rho }}{{\partial t}} = D\frac{{{\partial ^2}\rho }}{{\partial {y^2}}},\\
    D = 1.40 \times {10^{ - 5}}\;{{\rm m}^2}{{\rm s}^{ - 1}}
    \end{array}$} & \multirow{2}{*}{Clean dataset} & DHC-GEP      & $\frac{{\partial \rho }}{{\partial t}} = 1.0D\frac{{{\partial ^2}\rho }}{{\partial {y^2}}}$                & 0      \\
\cdashline{4-6}[1pt/1pt]
    &                      &                                & Original-GEP & $\frac{{\partial \rho }}{{\partial t}} = 1.40 \times {10^{ - 5}}\frac{{{\partial ^2}\rho }}{{\partial {y^2}}}$                & 0      \\ 
\cdashline{3-6}[1pt/1pt]
                                        &                      & \multirow{2}{*}{DSMC dataset}  & DHC-GEP      & $\frac{{\partial \rho }}{{\partial t}} = 0.984D\frac{{{\partial ^2}\rho }}{{\partial {y^2}}}$                & 0.027  \\
\cdashline{4-6}[1pt/1pt]
                                        &                      &                                & Original-GEP & $\frac{\partial \rho }{\partial t} =1.91\times 10^{-7}\left ( 72\frac{\partial^{2}  \rho }{\partial y^{2} }+72 \frac{\partial^{2}  \rho }{\partial y^{2} }/\frac{\partial  \rho }{\partial y } +\frac{1}{D^{2}\left ( D-\rho +1 \right ) } \right )  $                & 0.024  \\ 
\hdashline[1pt/1pt]
\multirow{4}{*}{Taylor-Green vortex}    & \multirow{4}{*}{$\begin{array}{l}
    \frac{{\partial {\omega _z}}}{{\partial t}} = \upsilon \left( {\frac{{{\partial ^2}{\omega _z}}}{{\partial {x^2}}} + \frac{{{\partial ^2}{\omega _z}}}{{\partial {y^2}}}} \right) - \left( {u\frac{{\partial {\omega _z}}}{{\partial x}} + v\frac{{\partial {\omega _z}}}{{\partial y}}} \right),\\
    \upsilon  = 1.17 \times {10^{ - 5}} \;{{\rm m}^2}{{\rm s}^{ - 1}}
    \end{array}$}   & \multirow{2}{*}{Clean dataset} & DHC-GEP      & $\frac{{\partial {\omega _z}}}{{\partial t}} = 1.0\upsilon \left( {\frac{{{\partial ^2}{\omega _z}}}{{\partial {x^2}}} + \frac{{{\partial ^2}{\omega _z}}}{{\partial {y^2}}}} \right)$                & 0      \\
\cdashline{4-6}[1pt/1pt]
    &                      &                                & Original-GEP & $\frac{{\partial {\omega _z}}}{{\partial t}} = 1.17 \times {10^{ - 5}}\left( {\frac{{{\partial ^2}{\omega _z}}}{{\partial {x^2}}} + \frac{{{\partial ^2}{\omega _z}}}{{\partial {y^2}}}} \right)$                & 0      \\ 
\cdashline{3-6}[1pt/1pt]
                                        &                      & \multirow{2}{*}{DSMC dataset}  & DHC-GEP      & $\frac{{\partial {\omega _z}}}{{\partial t}} = 0.981\upsilon \left( {\frac{{{\partial ^2}{\omega _z}}}{{\partial {x^2}}} + \frac{{{\partial ^2}{\omega _z}}}{{\partial {y^2}}}} \right)$                & 0.112  \\
\cdashline{4-6}[1pt/1pt]
                                        &                      &                                & Original-GEP & $\frac{{\partial {\omega _z}}}{{\partial t}}=-60.53\left ( \frac{\partial^{2}  u}{\partial y^{2} } /\left ( u+\frac{\partial v }{\partial x}+\frac{\partial \omega _{z}}{\partial y}+3\frac{\partial v}{\partial y}\frac{\partial^{2}  u}{\partial y^{2} } \frac{\partial^{2}  \omega _{z} }{\partial y^{2} }    \right ) +\frac{\partial v}{\partial x} \right ) $               & 0.096  \\ 
\hdashline[1pt/1pt]
Hybrid problem & $\frac{{\partial {\omega _z}}}{{\partial t}} = \upsilon \left( {\frac{{{\partial ^2}{\omega _z}}}{{\partial {x^2}}} + \frac{{{\partial ^2}{\omega _z}}}{{\partial {y^2}}}} \right) - \left( {u\frac{{\partial {\omega _z}}}{{\partial x}} + v\frac{{\partial {\omega _z}}}{{\partial y}}} \right),$  & Hybrid dataset  & DHC-GEP      & $\frac{{\partial {\omega _z}}}{{\partial t}} = 0.988\upsilon \left( {\frac{{{\partial ^2}{\omega _z}}}{{\partial {x^2}}} + \frac{{{\partial ^2}{\omega _z}}}{{\partial {y^2}}}} \right) - 0.988\left( {u\frac{{\partial {\omega _z}}}{{\partial x}} + v\frac{{\partial {\omega _z}}}{{\partial y}}} \right)$               & 0.082  \\
\hdashline[1pt/1pt]
\end{tabular}}
\caption{\label{tab2}Derived equations of DHC-GEP.}
\end{table}

Subsequently, we test DHC-GEP and Original-GEP on the DSMC dataset. All settings are consistent with those in the test on the clean dataset. The derived equations are shown in table \ref{tab2}. The coefficient of the equation discovered by DHC-GEP has a minor deviation from that of the theoretical equation, and the loss is not zero. This is because DSMC is a stochastic molecule-based method, the data of which are inherently noisy. It is impossible for DSMC to simulate a flow with the diffusion coefficient being exact $1.40 \times {10^{ - 5}}\;{{\rm m}^2}{{\rm s}^{ - 1}}$. Small fluctuations around the theoretical value are acceptable. In addition, calculating derivatives also introduces errors. Therefore, we believe that DHC-GEP has discovered the correct equation, while the derived equation of Original-GEP is clearly wrong for not sartisfying the fundamental dimensional homogeneity. Besides, we generate another dataset of diffusion flow with a diffusion coefficient being $3.0 \times {10^{ - 5}}\;{{\rm m}^2}{{\rm s}^{ - 1}}$ and calculate the losses of the two derived equations based on the new dataset. The losses for the derived equations of Original-GEP and DHC-GEP are 0.55 and 0.06, respectively. Therefore, the equation obtained by Original-GEP is clearly overfitting, while the equation obtained by DHC-GEP is generally applicable in different diffusion flows.

To test the sensitivity of DHC-GEP to hyperparameters, we conduct a parametric study on the length of head, the number of genes in a chromosome, and the number of individuals in a population. Specifically, the parameter spaces for these three hyperparameters are set to $\left \{ 10, 11, 12, 13, 14, 15\right \} $, $\left \{1, 2\right \}$ and $\left \{800, 1000, 1200, 1400, 1600, 1660\right \}$, respectively, resulting in a total of 72 ($6\times 2\times 6$) distinct parameter combinations. Other hyperparameters are kept unchanged, including the probabilities of invoking the genetic operators and the maximum number of evolution generations. In all 72 sets of parameter combinations, DHC-GEP consistently discovers the correct equation. In contrast, Original-GEP discovers 71 completely different results, all of which are overfitting and fail to satisfy the dimensional homogeneity constraint despite their lower losses than the correct equations.

Considering that DSMC is a simulation method at meso-scale, which coarse grains the molecular description to the hydrodynamic regime \citep{hadjiconstantinou2000analysis}, the influence of the coarse graining on the derived equations of DHC-GEP is discussed. For the same computational domain, we employ seven sets of sampling cells with different resolutions ($256\times 256$, $128\times 128$, $64\times 64$, $32\times 32$, $16\times 16$, $8\times 8$, $4\times 4$) to acquire macroscopic physical quantities. The derived equations based on these datasets are summarized in table \ref{coarse graining analysis}. It is encouraging to note that DHC-GEP is still capable of discovering the correct equation when the resolution is $16\times 16$ and the size of sampling cells is up to $6.3\lambda$. From the results, it can be concluded that DHC-GEP is robust to coarse graining, as long as the sampled data can accurately capture the gradients of macroscopic quantities.

\begin{table}
\centering
\renewcommand\arraystretch{2.4}
{
\begin{tabular}{cccc}
Resolution of sampling cells & Size of sampling cells & Derived equation & Loss   \\ 
\hdashline[1pt/1pt]
$256\times 256$                      & $0.4\lambda$           & $\frac{\partial \rho }{\partial t} =0.984D\frac{\partial^{2}  \rho }{\partial y^{2} }$                & 0.027  \\ 
\hdashline[1pt/1pt]
$128\times 128$                      & $0.8\lambda$           & $\frac{\partial \rho }{\partial t} =0.986D\frac{\partial^{2}  \rho }{\partial y^{2} }$                & 0.024  \\ 
\hdashline[1pt/1pt]
$64\times 64$                        & $1.6\lambda$           & $\frac{\partial \rho }{\partial t} =0.994D\frac{\partial^{2}  \rho }{\partial y^{2} }$                & 0.027  \\ 
\hdashline[1pt/1pt]
$32\times 32$                        & $3.1\lambda$           & $\frac{\partial \rho }{\partial t} =0.990D\frac{\partial^{2}  \rho }{\partial y^{2} }$                & 0.045  \\ 
\hdashline[1pt/1pt]
$16\times 16$                        & $6.3\lambda$           & $\frac{\partial \rho }{\partial t} =1.001D\frac{\partial^{2}  \rho }{\partial y^{2} }$                & 0.038  \\ 
\hdashline[1pt/1pt]
$8\times 8$                          & $12.5\lambda$          & $\frac{\partial \rho }{\partial t} =0.791\left ( D +\frac{\mu }{\rho } \right ) \frac{\partial^{2}  \rho }{\partial y^{2} }$                & 0.445  \\ 
\hdashline[1pt/1pt]
$4\times 4$                          & $25\lambda$            & $\frac{\partial \rho }{\partial t} =-0.075\left ( D\frac{\partial \rho }{\partial y}\frac{\partial^{2}  \rho }{\partial y^{2} }+\left ( \mu +D\rho  \right ) \frac{\partial^{3}  \rho }{\partial y^{3} }\right )/\frac{\partial  \rho }{\partial y }  $                & 0.516  \\ 
\hdashline[1pt/1pt]
\multicolumn{2}{c}{Theoretical equation}              & \multicolumn{2}{c}{$\frac{\partial \rho }{\partial t} =D\frac{\partial^{2}  \rho }{\partial y^{2} }$}    
\end{tabular}}
\caption{\label{coarse graining analysis}Derived equations of DHC-GEP based on the datasets with different resolutions.}
\end{table}

In DHC-GEP, the overfitting results are automatically filtered out due to not satisfying the dimensional homogeneity. On the contrary, Original-GEP always favours the equations with smaller loss, so it tends to converge to overfitting results if data are noisy. Our numerical experiences show that Original-GEP may discover the correct equations only when tuning the hyperparameters repeatedly and terminating the evolution at a mediate generation (when overfitting equations have not appeared).

We compare the computational cost needed per 1 000 generations of evolution. Based on the 3.5 GHz Intel Xeon E5-1620 processor, the average CPU runtime of Original-GEP (443 seconds) is almost twice that of DHC-GEP (216 seconds). The main reason is that DHC-GEP can identify some individuals as invalid through dimensional verification and then skip the process of evaluating losses for these individuals. It can still save computational time despite the extra expense of dimensional verification. Additionally, as the complexity of problem increases, the number of invalid individuals also increases, so the advantage of the computational efficiency of DHC-GEP becomes more significant.

Furthermore, we conduct a much more challenging test in which we do not include the transport coefficients in the terminal set. Specifically, the terminal set is $\left\{ {\tau  ,\lambda ,\frac{{\partial \rho }}{{\partial y}},\frac{{{\partial ^2}\rho }}{{\partial {y^2}}}{\rm{,}}\frac{{{\partial ^3}\rho }}{{\partial {y^3}}}} \right\}$, which involves only the fundamental physical property parameters of gas (molecular mean collision time ($\tau $) and molecular mean free path ($\lambda $)), but excludes the diffusion coefficient ($D$). Other hyperparameters are the same as those in the original test. The derived equation is
\begin{equation}
\frac{{\partial \rho }}{{\partial t}} = 0.579\frac{\lambda ^{2} }{\tau }\frac{{{\partial ^2}\rho }}{{\partial {y^2}}}.
\label{diffusion equation without diffusion coefficient}
\end{equation}
According to (\ref{diffusion coefficient}), the coefficient ($0.579\frac{\lambda ^{2} }{\tau }$) is essentially the diffusion coefficient. This result demonstrates that DHC-GEP has the capability to derive accurate equations with minor prior knowledge of the essential transport coefficients, and can deduce the essential transport coefficients from fundamental physical property parameters using dimensional homogeneity as a constraint.

\subsection {Vorticity transport equation}

\begin{figure}  
\centering  
\subfloat{\includegraphics[height=5cm]{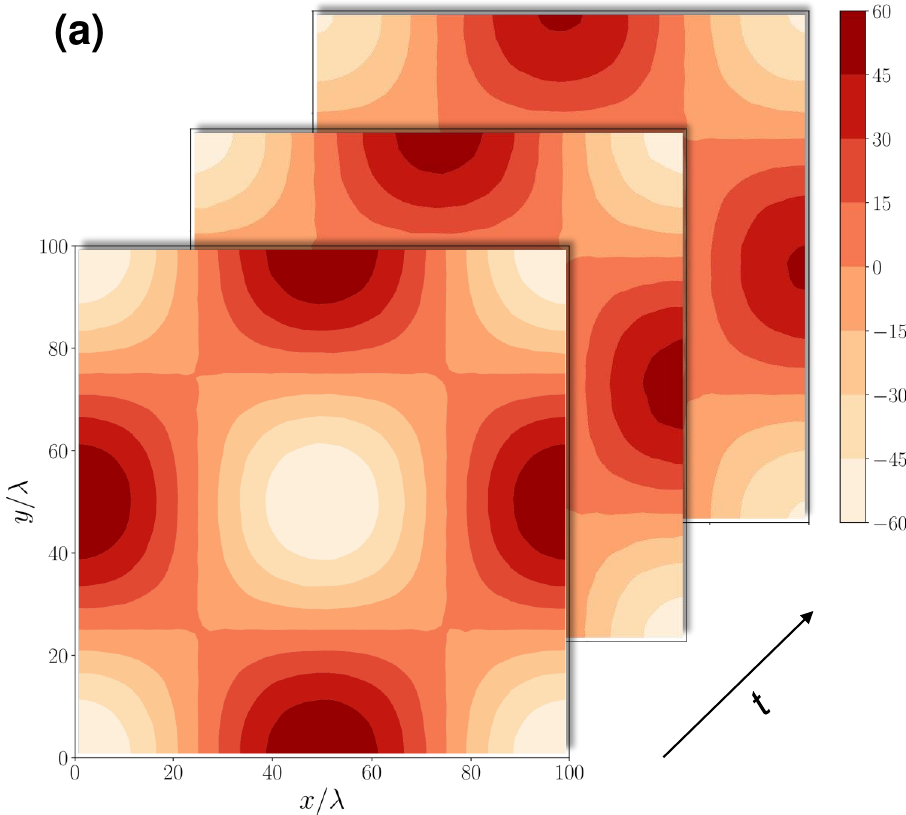}}
\hspace{0.1cm} 
\subfloat{\includegraphics[height=5cm]{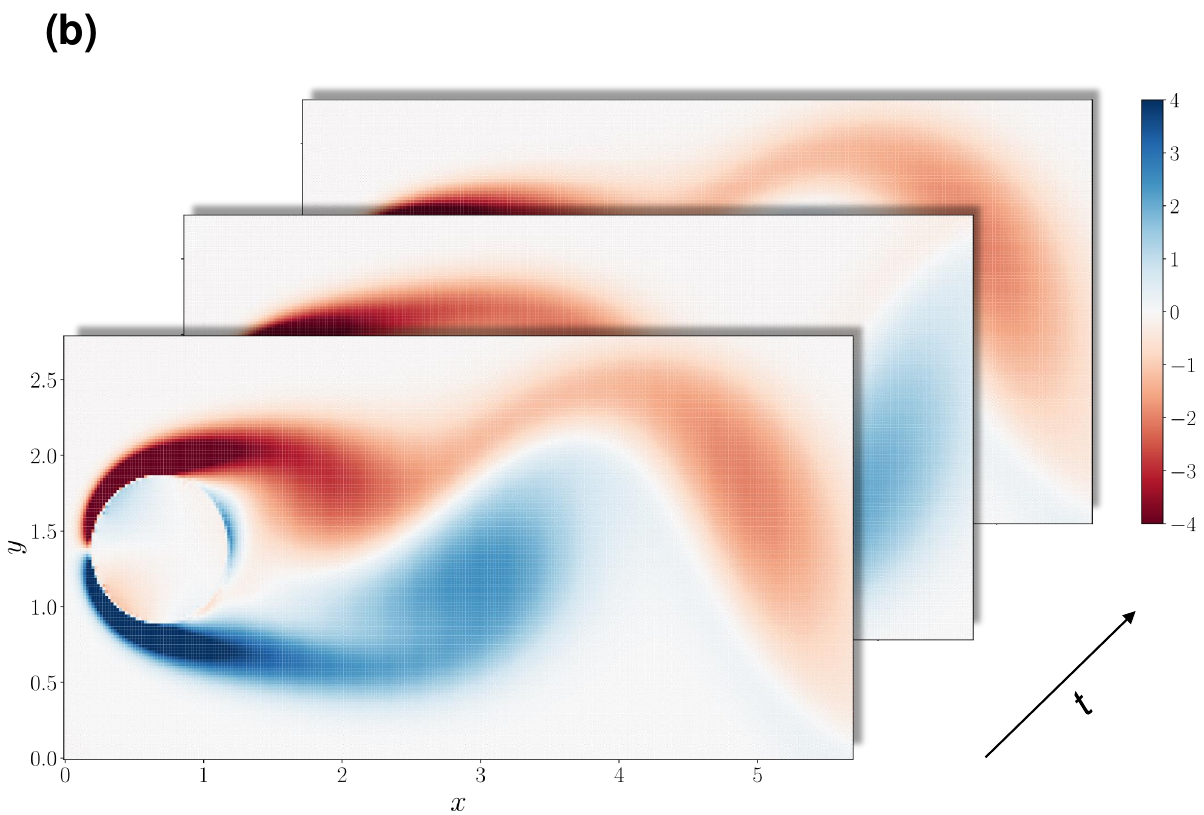}}
\caption{Temporal evolution of vorticity. (a) Taylor-Green vortex at the instants of $t = 0,100\tau ,\;200\tau .$ Here, $\tau$ and $\lambda$ are the mean collision time and mean free path of argon gas molecules at standard condition, respectively. (b) Viscous flow around a cylinder at the instants of $t = 0\;{\rm s},3\;{\rm s},\;6\;{\rm s}.$} 
\label{fig3}
\end{figure}

We employ DSMC to simulate the temporal evolution of Taylor-Green vortex, as shown in figure \ref{fig3}(a). The simulation model is approximately the same as that of diffusion flow, including simulation conditions, geometry, and boundary conditions. The major difference is that the initial distribution of macroscopic velocity is Taylor-Green vortex,
\begin{equation}
    \begin{dcases} {\begin{array}{*{20}{l}}
        {u = {v_0}\cos \left( x \right)\sin \left( y \right)}\vspace{0.8ex}\\
        {v =  - {v_0}\sin \left( x \right)\cos \left( y \right)}
        \end{array}} \end{dcases},
\label{su_eq1}
\end{equation}
where $u$ and $v$ are the velocities in the horizontal ($x$) and vertical ($y$) directions, respectively, and ${v_0}$ is the initial amplitude of velocity and equals $30\;{\rm m} \cdot {{\rm s}^{ - 1}}$ in our simulation.

The computational cell and time step are set the same as those in the diffusion flow. Additionally, we introduce sampling cells. In this case, each sampling cell consists of 16 computational cells, i.e., the whole computational domain has $64 \times 64$ sampling cells. The macroscopic physical quantities obtained with sampling cells are cleaner than those obtained with computational cells, due to more molecules being sampled in each cell. In addition, to further reduce the statistical errors, 10 independent simulations are conducted with different random number sequences to make an ensemble average. During simulation, we sample the macroscopic velocities ($u,v$) and vorticities (${\omega _z}$) at the instants of $t = 0,\;10\tau ,\;20\tau ,...,\;310\tau $. The final dataset is composed of 131 072 ($64 \times 64 \times 32$) data points distributed in $64 \times 64$ sampling cells at 32 instants.

The theoretical governing equation is the vorticity transport equation,
\begin{equation}
\frac{{\partial {\omega _z}}}{{\partial t}} = \upsilon \left( {\frac{{{\partial ^2}{\omega _z}}}{{\partial {x^2}}} + \frac{{{\partial ^2}{\omega _z}}}{{\partial {y^2}}}} \right) - \left( {u\frac{{\partial {\omega _z}}}{{\partial x}} + v\frac{{\partial {\omega _z}}}{{\partial y}}} \right).
\label{eq3}
\end{equation}
Here, ${\omega _z}$ is the vorticity in the $z$ direction. $\upsilon $ is the kinematic viscosity and approximately equals $1.17 \times {10^{ - 5}}\;{{\rm m}^2}{{\rm s}^{ - 1}}$. Compared with the diffusion equation (\ref{eq2}), this equation is more complex, involving multiple variables and nonlinear terms. 

In this case, we define the target variable as $\frac{{\partial {\omega _z}}}{{\partial t}}$, the function set as $\left\{ { + , - , \times , \div } \right\}$, and the terminal set as $\left\{ {\upsilon ,u,v,\omega _z,\frac{{\partial u}}{{\partial x}},\frac{{{\partial ^2}u}}{{\partial {x^2}}},\frac{{\partial u}}{{\partial y}},\frac{{{\partial ^2}u}}{{\partial {y^2}}},\frac{{\partial v}}{{\partial x}},\frac{{{\partial ^2}v}}{{\partial {x^2}}},\frac{{\partial v}}{{\partial y}},\frac{{{\partial ^2}v}}{{\partial {y^2}}},\frac{{\partial \omega _z }}{{\partial x}},\frac{{{\partial ^2}\omega _z }}{{\partial {x^2}}},\frac{{\partial \omega _z }}{{\partial y}},\frac{{{\partial ^2}\omega _z }}{{\partial {y^2}}}} \right\}$.

We also test the performances of DHC-GEP on the clean dataset and the DSMC dataset sequentially. The results are provided in table \ref{tab2}. An interesting observation is that the derived equations of DHC-GEP miss two convective terms $\left( {u\frac{{\partial {\omega _z}}}{{\partial x}},v\frac{{\partial {\omega _z}}}{{\partial y}}} \right)$. This is caused by the unique feature of the Taylor-Green vortex. The analytical solution of the Taylor-Green vortex is
\begin{equation}
    \begin{dcases}\begin{array}{*{35}{l}}
        u=v_{0}\cdot {\rm cos}\left (x \right ){\rm sin}\left (y \right ) {\exp \left( { - 2\upsilon t} \right)} \vspace{0.8ex}\\
        v=-v_{0}\cdot {\rm sin}\left (x \right ){\rm cos}\left (y \right ) {\exp \left( { - 2\upsilon t} \right)}\vspace{0.8ex}\\
        {{\omega _z}} =-2v_{0}\cdot {\rm cos}\left (x \right ){\rm cos}\left (y \right ) {\exp \left( { - 2\upsilon t} \right)}
    \end{array}\end{dcases}.
\label{eq4}
\end{equation}
Substituting (\ref{eq4}) into the convective terms $\left( {u\frac{{\partial {\omega _z}}}{{\partial x}},v\frac{{\partial {\omega _z}}}{{\partial y}}} \right)$, it is clear that the sum of the convective terms is automatically zero. Therefore, the equations without convective terms are also correct for the Taylor-Green vortex. It can be considered as a specialized variant of the complete vorticity transport equation in the Taylor-Green vortex.

To discover the complete vorticity transport equation and validate the performance of DHC-GEP on discovering nonlinear terms, we further consider a viscous flow around a cylinder at Reynolds number being 100, as shown in figure \ref{fig3}(b). The dataset is the open access dataset provided by \citet{rudy2017data}, and the theoretical governing equation is 
\begin{equation}
\frac{{\partial {\omega _z}}}{{\partial t}} = 0.01 \cdot \left( {\frac{{{\partial ^2}{\omega _z}}}{{\partial {x^2}}} + \frac{{{\partial ^2}{\omega _z}}}{{\partial {y^2}}}} \right) - \left( {u\frac{{\partial {\omega _z}}}{{\partial x}} + v\frac{{\partial {\omega _z}}}{{\partial y}}} \right).
\label{eq5}
\end{equation}
For the sake of simplicity, we regard the data as dimensional data with kinematic viscosity $\upsilon = 0.01\;{{\rm m}^2}{{\rm s}^{ - 1}}$. We randomly sampled 200 spatiotemporal data points from the Taylor-Green vortex dataset and another 200 spatiotemporal data points from the flow around cylinder dataset, forming a hybrid dataset consisting of 400 spatiotemporal data points. The hyperparameters of DHC-GEP are consistent with those in the case of the Taylor-Green vortex. The results are provided in table \ref{tab2}. DHC-GEP discovers the correct and complete vorticity transport equation. This demonstrates that in some special cases, where the overfitting equations may satisfy the dimensional homogeneity and fit well with the training data, employing multiple datasets can help mitigate overfitting. Overfitting equations are generally dataset-specific. While an overfitting equation may adequately describe a particular dataset, it cannot be expected to generalize well to multiple datasets. Besides, note that the training dataset consisting of 400 spatiotemporal data points is relatively small. Generally, if the size of dataset is small, the information carried by data is sparse, leading to the derived equations overfitting to specific phenomena. However, in DHC-GEP, the overfitting equations are automatically filtered out by the constraint of dimensional homogeneity. 

\section{Application on discovering unknown constitutive relations}
\label{sec:AODUCR}

The general governing equations for fluid flows are the conservation equations of mass, momentum and energy as follows:
\begin{equation}
    \begin{dcases}\begin{array}{*{35}{l}}
        \frac{{D\rho }}{{Dt}} + \rho \frac{{\partial {v_k}}}{{\partial {x_k}}} = 0, \vspace{0.8ex}\\
        \rho \frac{{D{v_i}}}{{Dt}} + \frac{{\partial p}}{{\partial {x_i}}} + \frac{{\partial {\tau _{ik}}}}{{\partial {x_k}}} = \rho {F_i},\vspace{0.8ex}\\
        \frac{3}{2}\frac{{D\theta }}{{Dt}} + \frac{{\partial {q_k}}}{{\partial {x_k}}} =  - \left( {p{\delta _{ij}} + {\tau _{ij}}} \right)\frac{{\partial {v_i}}}{{\partial {x_j}}}.
    \end{array}\end{dcases}.
\label{eq6}
\end{equation}
Here, $\frac{D}{{Dt}}{\rm{=}}\frac{\partial }{{\partial t}}{\rm{ + }}{v_i}\frac{\partial }{{\partial {x_i}}}$ is the substantial derivative, and $\theta  = {k}_{B}T/m$ is the temperature in $energy/mass$ unit. Besides the five basic physical variables $\rho$, ${v_i}$ and $\theta$, there are eight additional variables, i.e., the viscous stress ${\tau _{ij}}$ and heat flux ${q_i}$. To numerically solve (\ref{eq6}), the additional constitutive relations that close the viscous stress and heat flux are needed. In the continuum regime, the NSF equations are widely employed, which assumes that the viscous stress/heat flux is linearly proportional to the local strain rate/temperature gradient. However, for strong non-equilibrium flows, the NSF equations are not valid \citep{boyd1995predicting}. 

Alternatively, high-order constitutive relations have been derived from the Boltzmann equation using the Chapman-Enskog method \citep{chapman1990mathematical}, including Burnett equations \citep{burnett1936distribution}, super-Burnett equations \citep{shavaliyev1993super} and augmented-Burnett equations \citep{zhong1993stabilization}, to account for the non-equilibrium effects. However, despite being proven to be superior to NSF equations, these equations are still unsatisfactory in strong non-equilibrium flows. In this work, we employ DHC-GEP to discover the unknown constitutive relations in two representative non-equilibrium flows, as the examples to illustrate how to apply DHC-GEP to discover unknown governing equations.

To avoid any potential misleading, it is important to emphasize that we do not employ DHC-GEP to discover the complex governing equations like equation (\ref{eq6}). In this work, we are focused on discovering the constitutive relations for the two unclosed variables within equation (\ref{eq6}), i.e., the viscous stress and heat flux.

\subsection {One-dimensional shock wave}

\begin{figure}  
\centering  
\includegraphics[width=5cm]{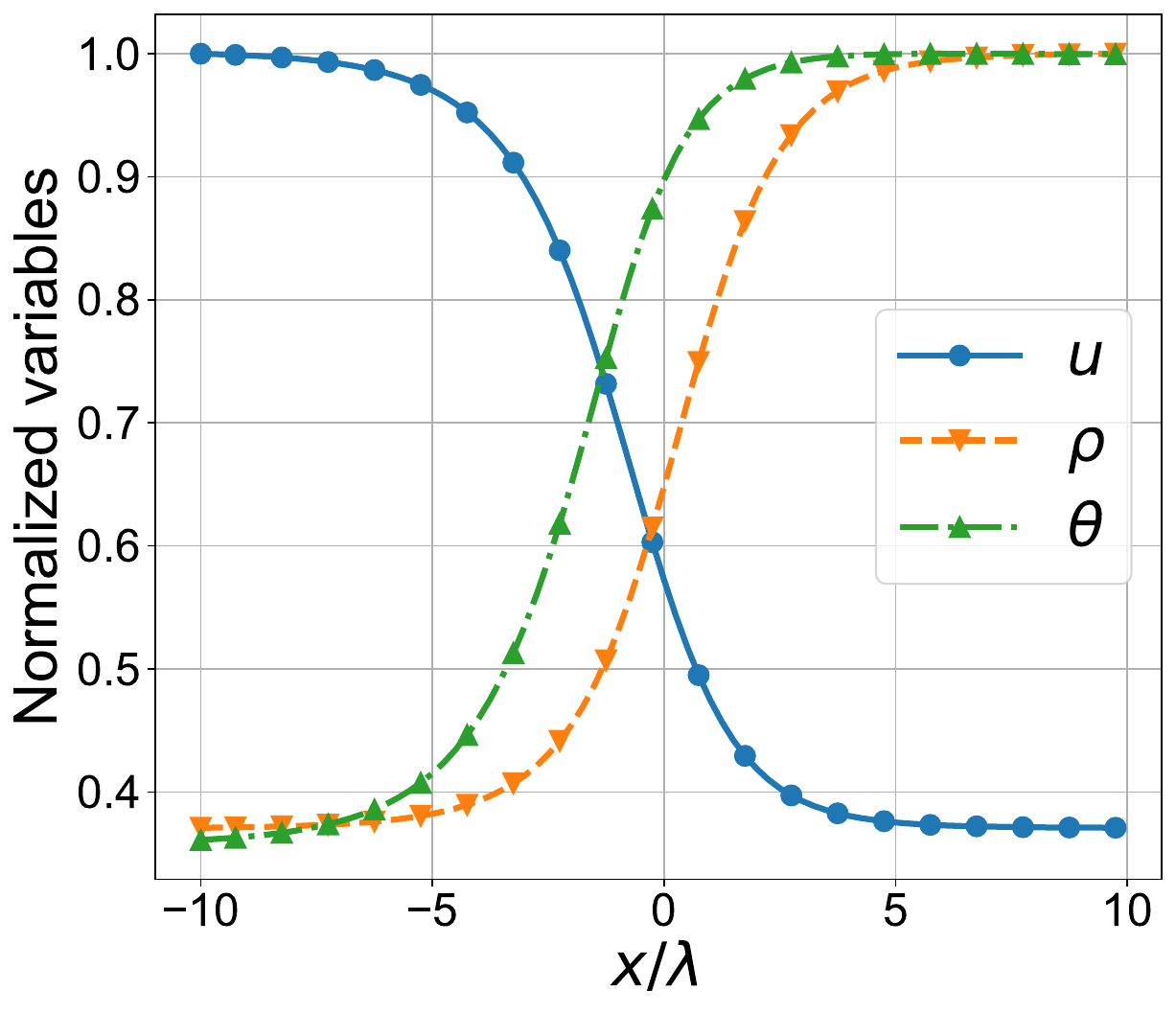}
\caption{General structure of one-dimensional shock. Each variable is normalized with its maximum value in the computational domain.} 
\label{fig4}
\end{figure}

In the research community of non-equilibrium flows, one-dimensional shock wave is a benchmark to validate solvers and formulations in numerical computations. We employ DSMC to simulate the one-dimensional shock wave of argon gas, the general structure of which is shown in figure \ref{fig4}. 

\begin{figure}  
\centering  
\includegraphics[width=9.5cm]{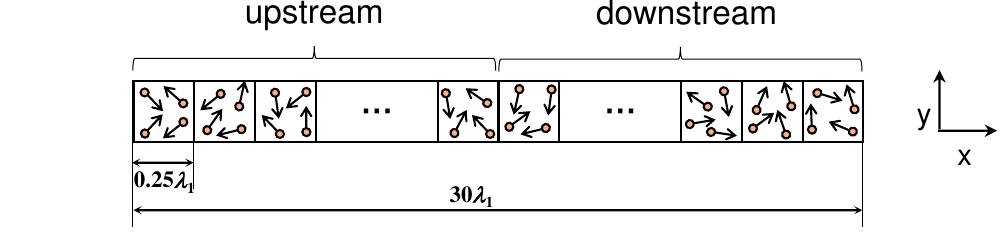}
\caption{Schematic diagram of the computational domain for one-dimensional shock.} 
\label{fig_mesh_1_D_shock}
\end{figure}

The one-dimensional shock essentially connects two equilibrium states. In this work, for upstream (i.e., the free stream), the temperature (${T_1}$) is $300\;{\rm K}$, the density (${\rho _1}$) is $1.067 \times {10^{ - 4}}\;{\rm kg} \cdot {{\rm m}^{ - 3}}$, and the mean free path (${\lambda _1}$) is $1.114 \times {10^{ - 3}}\;{\rm m}$. For downstream, the state variables are computed with Rankine-Hugoniot relations
\begin{equation}
    \begin{dcases}\begin{array}{*{35}{l}}
        {T_2} = \frac{{\left( {5Ma_\infty ^2 - 1} \right)\left( {Ma_\infty ^2 + 3} \right)}}{{16Ma_\infty ^2}}{T_1}\vspace{1.2ex}\\
        {\rho _2} = \frac{{4Ma_\infty ^2}}{{Ma_\infty ^2 + 3}}{\rho _1}\vspace{1.2ex}\\
        {v_2} = \frac{{Ma_\infty ^2 + 3}}{{4Ma_\infty ^2}}{v_1}
    \end{array}\end{dcases}.
\label{su_eq2}
\end{equation}
Here, $M{a_\infty }$ and ${v_1}$ are the Mach number and velocity of the free stream, respectively. Note that for this one-dimensional problem, the velocity direction of each position in the flow field is along the $x$ direction, and all the flow variables vary along only the $x$ direction. Therefore, we set 1 cell along the $y$ and $z$ directions, but 120 cells along the $x$ direction. A schematic diagram is shown in figure \ref{fig_mesh_1_D_shock}. The lengths of the computational domain and computational cell are $30{\lambda _1}$ and $0.25{\lambda _1}$, respectively. The time step is $0.1\tau $.

During simulation, we sample the macroscopic velocities ($u$), densities ($\rho$), temperatures ($T$), and heat flux in the $x$ direction (${q_x}$) at the instants of $t = 10\;000\tau ,\;10\;010\tau ,\;10\;020\tau ,...,$ $50\;000\tau $. Then, we obtain the final data by averaging all the sample data.

In this case, the target variable is defined as the heat flux in the $x$ direction (${q_x}$). The function set is defined as $\left\{ { + , - , \times , \div } \right\}$ as usual, while the terminal set is meticulously constructed to embed physics knowledge as follows.
\begin{itemize}
    \item Constitutive relations describe the local transport mechanisms of momentum and energy, and hence we regard local non-equilibrium parameters as key factors. Specifically, the gradient-length local (GLL) Knudsen number defined as $K{n_Q} = \frac{{{\lambda _l}}}{Q}\frac{\partial Q}{\partial x} $ is selected. The local non-equilibrium characteristics intensify as the increase of the absolute value of $K{n_Q}$. Here, ${\lambda _l}$ is the local mean free path, and $Q$ represents state variables, including temperature ($\theta $) and density ($\rho$).
    \item The transports of momentum and energy are driven by the gradients of state variables. As a result, the gradient terms are selected, including $\frac{{\partial u}}{{\partial x}},\frac{{\partial \rho }}{{\partial x}},\frac{{\partial \theta }}{{\partial x}},\frac{{{\partial ^2}u}}{{\partial {x^2}}},\frac{{{\partial ^2}\rho }}{{\partial {x^2}}},\frac{{{\partial ^2}\theta }}{{\partial {x^2}}},\frac{{{\partial ^3}u}}{{\partial {x^3}}},\frac{{{\partial ^3}\rho }}{{\partial {x^3}}}$, and $\frac{{{\partial ^3}\theta }}{{\partial {x^3}}}$.
    \item The state variables themselves are important factors. Besides, it is noteworthy that the constitutive relations should satisfy Galilean invariance \citep{han2019uniformly,li2021learning}, which means that the constitutive relations cannot contain velocity ($u$) explicitly outside the partial differential operators. The proof is provided in Appendix \ref{appC}. Therefore, the state variables ($\theta$ and $\rho$) excluding velocity are selected. This is also the reason that we do not select the GLL Knudsen number of velocity.
    \item The parameters representing the physical properties of gas are selected, including local viscosity ($\mu$), local heat conductivity ($\kappa $), heat capacity ratio ($\gamma  = 5/3$ for argon gas), and viscosity exponent ($\omega  = 1$ for Maxwell molecules).
\end{itemize}

Finally, the terminal set is $\{K{n_\theta },K{n_\rho },\frac{{\partial u}}{{\partial x}},\frac{{\partial \rho }}{{\partial x}},\frac{{\partial \theta }}{{\partial x}},\frac{{{\partial ^2}u}}{{\partial {x^2}}},\frac{{{\partial ^2}\rho }}{{\partial {x^2}}},\frac{{{\partial ^2}\theta }}{{\partial {x^2}}},\frac{{{\partial ^3}u}}{{\partial {x^3}}},\frac{{{\partial ^3}\rho }}{{\partial {x^3}}},\frac{{{\partial ^3}\theta }}{{\partial {x^3}}},\rho ,\theta ,\mu ,\kappa,$ $\gamma ,\omega \} $.

Moreover, constitutive relations are ought to satisfy the second law of thermodynamics. Theoretically, whether a constitutive relation satisfies the second law of thermodynamics can be determined through the Clausius-Duhem inequality \citep{comeaux1995evaluation}
\begin{equation}
    \rho \frac{{D{s_{eq}}}}{{Dt}} + \nabla  \cdot \frac{\mathbf{q}}{\theta } =  - \frac{1}{\theta }\mathbf{\tau} :\nabla \mathbf{v} - \frac{1}{{{\theta ^2}}}\mathbf{q} \cdot \nabla \theta  \ge 0.
\label{eq9}
\end{equation}
The two terms on the left-hand side are the local increase rate of entropy and the reversible outflow of entropy, respectively. The sum of the two terms on the right-hand side is the total entropy production. The second law of thermodynamics demands that the total entropy production must be non-negative. Note that the total entropy production is the sum of sub-entropy productions of all high-order macroscopic variables (the viscous stress ${\tau _{ij}}$ and heat flux ${q_i}$). If each sub-entropy production is non-negative, the total entropy production is naturally non-negative. This work deals with the discovery of the constitutive equation of a single high-order macroscopic variable, rather than all of them. Hence, the entropy production discussed in this work refers to the contribution of a single high-order macroscopic variable's constitutive relation. Specifically, in the one-dimensional shock wave case, the target variable is the heat flux in the $x$ direction (${q_x}$), and thus the entropy production considered here is
\begin{equation}
    S_{p,q_{x} } =  - \frac{1}{{{\theta ^2}}}q_{x}\frac{\partial \theta }{\partial x}.
\label{entropy_q_x}
\end{equation}

To ensure that the resulting constitutive equation satisfies the second law of thermodynamics, we introduce this law as a constraint into the loss function (\ref{eq16}). Specifically, we incorporate the constraint in a soft form by adding a loss term ($\mathrm{L}_{S_p}$) that represents the ratio of the number of the data points with negative entropy production to the total number of data points. The modified loss function is
\begin{equation}
    \mathrm{Loss}=\mathrm{L}_{\mathrm{MRE}}+\alpha \mathrm{L}_{S_p}=\frac{1}{N} \sum_{i=1}^{N}\left | \frac{\widehat{Y}_i-Y_i}{Y_i}\right |+\alpha\frac{\left \langle S_p \right \rangle _{-} }{N}.
\label{modified_loss}
\end{equation}
Here, the variable with a superscript $\wedge$ is the predicted variable, and $N$ is the total number of data points. $\left \langle S_p \right \rangle _{-}$ is the operator that counts the number of data points with negative entropy production, and $\alpha$ is the weighting factor that controls the importance of the constraint. 

We employ the DHC-GEP, combined with the constraint of the second law of thermodynamics, to discover the underlying constitutive relation based on the data of two cases with the freestream Mach number $M{a_\infty } = 3.0$ and $4.0$. We sequentially set $\alpha$ to 0.01, 0.1, 0.2, 0.3, 0.4, 0.5, 0.6, 0.7, 0.8, 0.9, and 1.0. Each $\alpha$ corresponds to an independent training run and a resulting equation with relatively low $\mathrm{L}_{\mathrm{MRE}}$. Among them, the equation with the smallest $\mathrm{L}_{S_p}$ is ultimately selected as the optimal equation. For this case, the optimal equation is 
\begin{equation}
    {q_x} =  - 0.468\left( {6K{n_\rho }+\frac{{K{n_\theta }}}{{K{n_\rho }}} + \frac{{K{n_\rho }}}{{K{n_\theta }}}} \right)\kappa \frac{{\partial \theta }}{{\partial x}},
\label{eq7}
\end{equation}
which is derived with $\alpha=0.3$. The entropy production is
\begin{equation}
    S_{p,q_{x} } =  \frac{0.468\kappa}{\theta ^{2} } \left( {6K{n_\rho } +\frac{{K{n_\theta }}}{{K{n_\rho }}} + \frac{{K{n_\rho }}}{{K{n_\theta }}}} \right) \left ( \frac{{\partial \theta }}{{\partial x}} \right ) ^{2} .
\label{entropy_q_x_shock}
\end{equation}
According to figure \ref{fig4}, for the one-dimensional shock wave case, the gradients of density and temperature are non-negative. Additionally, according to the definition of the local Knudsen number ($K{n_Q} = \frac{{{\lambda _l}}}{Q}\frac{\partial Q}{\partial x} $), both $K{n_\rho }$ and $K{n_\theta }$ are thus non-negative. Therefore, from a mathematical perspective, the entropy production (\ref{entropy_q_x_shock}) must be non-negative. As a comparison, whether the Burnett equation, super-Burnett equation and augmented-Burnett equation satisfy the second law of thermodynamics is obscure \citep{comeaux1995evaluation}.

\begin{figure}  
\centering 
\includegraphics[width=13.7cm]{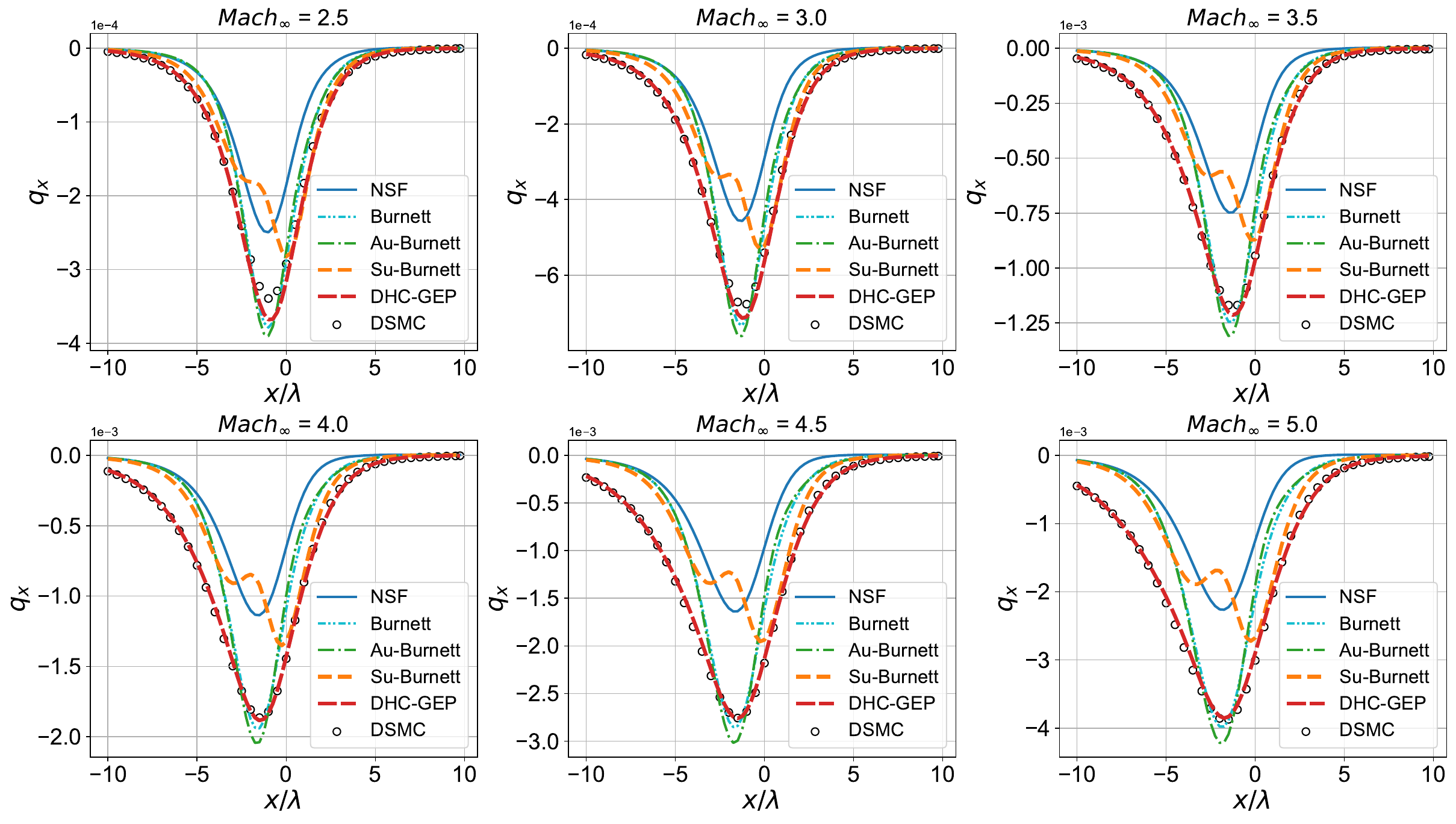}
\caption{Comparisons of the NSF equation, Burnett equation, augmented-Burnett equation (Au-Burnett), super-Burnett equation (Su-Burnett), derived equation of DHC-GEP, and DSMC (exact solution) at $M{a_\infty } = 2.5,\;3.0,\;3.5,\;4.0,\;4.5$ and $5.0$ for the one-dimensional shock wave case.}
\label{fig5}
\end{figure}

We compare the results predicted by (\ref{eq7}) with those predicted by the NSF equation, Burnett equation, augmented-Burnett equation, and super-Burnett equation in figure \ref{fig5}. For a quantitative comparison, we define the relative error as 
\begin{equation}
    \mathrm{error} = \frac{{\sqrt {\sum\limits_{i = 1}^N {{{\left( {{{\hat q}_{x,i}} - {q_{x,i}}} \right)}^2}} } }}{{\sqrt {\sum\limits_{i = 1}^N {q_{x,i}^2} } }}.
\label{eq8}
\end{equation}
The relative errors for each constitutive relation are summarized in table \ref{tab3}. It can be found that the derived constitutive relation of DHC-GEP exhibits higher accuracy than other equations over a wide range of $M{a_\infty}$, and is applicable to cases beyond the parameter space of the training data, i.e., $2.0 \le \; M{a_\infty} < 3.0$ and $4.0 < M{a_\infty} \le \; 5.5$.

Furthermore, to investigate the sensitivity to hyperparameters in non-equilibrium cases, we conduct a parametric study. Specifically, we focus on the sensitivity to the length of head, number of genes in a chromosome, and number of individuals in a population. The parameter spaces for these three hyperparameters are set to $\left \{ 10, 11, 12, 13, 14, 15\right \} $, $\left \{1, 2\right \}$ and $\left \{800, 1000, 1200, 1400, 1600, 1660\right \}$, respectively, resulting in a total of 72 ($6\times 2\times 6$) distinct hyperparameter combinations. Other hyperparameters, including the probability of invoking the genetic operators, are kept unchanged. Among the total of 72 distinct hyperparameter combinations, DHC-GEP successfully derives the same equation as (\ref{eq7}) based on 38 of these hyperparameter combinations. In contrast, Original-GEP discovers 70 completely different results, all of which fail to satisfy even the fundamental dimensional homogeneity constraint, let alone other physical constraints. It can be concluded that in non-equilibrium cases, while DHC-GEP is not entirely insensitive to model hyperparameters, it does exhibit significantly reduced sensitivity compared to Original-GEP.

\begin{table}
\centering
\resizebox{\textwidth}{8mm}{
\begin{tabular}{lccccccccccc} 
            & \multicolumn{1}{l}{$M{a_\infty } = 2.0$} & \multicolumn{1}{l}{$M{a_\infty } = 2.3$} & \multicolumn{1}{l}{$M{a_\infty } = 2.5$} & \multicolumn{1}{l}{$M{a_\infty } = 3.0$} & \multicolumn{1}{l}{$M{a_\infty } = 3.2$} & \multicolumn{1}{l}{$M{a_\infty } = 3.5$} & \multicolumn{1}{l}{$M{a_\infty } = 4.0$} & \multicolumn{1}{l}{$M{a_\infty } = 4.2$} & \multicolumn{1}{l}{$M{a_\infty } = 4.5$} & \multicolumn{1}{l}{$M{a_\infty } = 5.0$} & \multicolumn{1}{l}{$M{a_\infty } = 5.5$}   \\ 
DHC-GEP     & 0.101   & 0.093   & 0.086   & 0.057   & 0.053   & 0.045   & 0.041   & 0.044   & 0.046   & 0.050   & 0.055    \\
NSF         & 0.251   & 0.321   & 0.357   & 0.431   & 0.451   & 0.477   & 0.509   & 0.518   & 0.530   & 0.545   & 0.555    \\
Burnett     & 0.120   & 0.150   & 0.164   & 0.204   & 0.219   & 0.238   & 0.266   & 0.274   & 0.284   & 0.298   & 0.306    \\
Au-Burnett & 0.127   & 0.164   & 0.180   & 0.221   & 0.237   & 0.257   & 0.284   & 0.293   & 0.303   & 0.316   & 0.325    \\
Su-Burnett & 0.206   & 0.255   & 0.282   & 0.333   & 0.349   & 0.370   & 0.396   & 0.406   & 0.415   & 0.428   & 0.437    \\
\end{tabular}}
\caption{\label{tab3}Relative errors of the derived equation of DHC-GEP, NSF equation, Burnett equation, augmented-Burnett equation (Au-Burnett), and super-Burnett equation (Su-Burnett) for the one-dimensional shock wave case.}
\end{table}

\subsection {Rarefied Poiseuille flow}

\begin{figure}  
\centering  
\subfloat{\includegraphics[width=4.6cm]{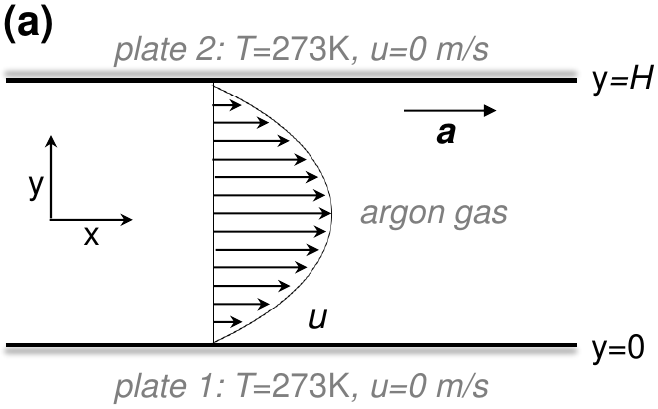}}
\hspace{1cm} 
\subfloat{\includegraphics[width=5.5cm]{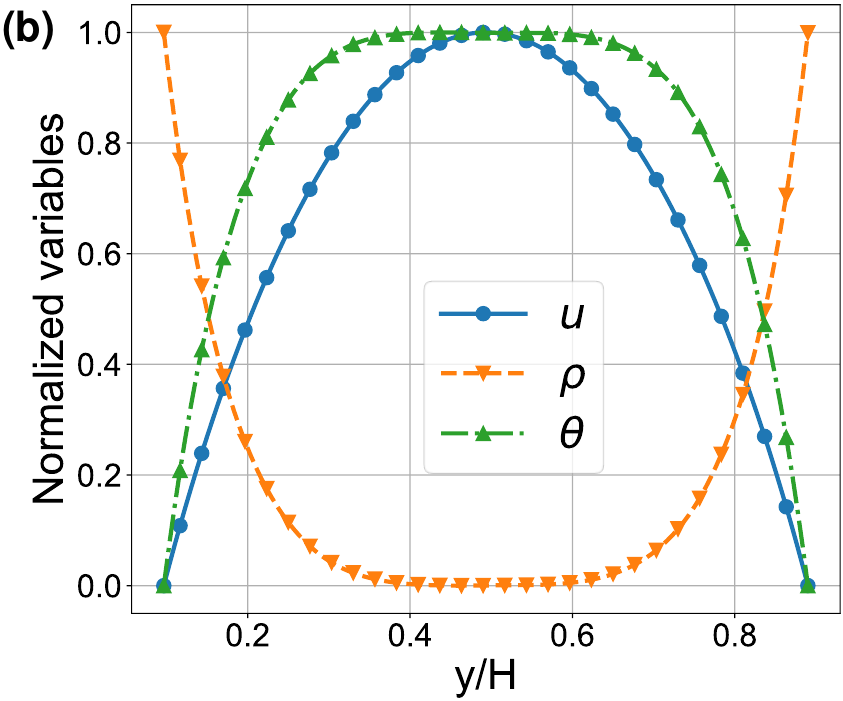}}
\caption{(a) Schematic diagram of Poiseuille flow. (b) Profiles of velocity, density and temperature along the direction normal to the plate. Each variable ($U$) is normalized with ${U_{{\rm Normal}}} = \left( {U - {U_{\min }}} \right)/\left( {{U_{\max }} - {U_{\min }}} \right)$.} 
\label{fig6}
\end{figure}

Poiseuille flow is a flow confined between two infinite, parallel and relative static plates. The gas is driven by an external force in the $x$ direction. The external force is uniformly distributed along the $y$ direction. A schematic diagram of Poiseuille flow is shown in figure \ref{fig6}. The global Knudsen number ($Kn$) is defined as 
\begin{equation}
    Kn = \frac{\lambda }{H}.
\label{su_eq3}
\end{equation}
Here, $H$ is the distance between two plates, and $\lambda$ is the mean free path of argon gas molecules at standard condition. For different $Kn$, $\lambda$ remains the same, while $H$ changes according to (\ref{su_eq3}). The uniform external force is implemented through a uniform acceleration in molecular simulations. The accelerations for different $Kn$ are listed in table \ref{su_tab1}.

\begin{table}
\centering
\begin{tabular}{ccccccccc} 
                                                            & $Kn=0.01$ & $Kn=0.1$ & $Kn=0.2$ & $Kn=0.3$ & $Kn=0.35$ & $Kn=0.4$ & $Kn=0.5$ & $Kn=0.6$  \\ 
\begin{tabular}[c]{@{}c@{}}Acceleration\\($ \times {10^9}\;{\rm m} \cdot {{\rm s}^{ - 2}}$)\end{tabular} & 3.31    & 320    & 444    & 716    & 856     & 986    & 1 240   & 1 490    \\
\end{tabular}
\caption{\label{su_tab1}Accelerations for different $Kn$.}
\end{table}

All the flow variables vary along only the $y$ direction. Therefore, we set 1 cell along the $x$ and $z$ directions, but 3 000 cells along the $y$ direction. The time step is $0.2\tau $. During simulation, we sample the macroscopic velocities ($u$), densities ($\rho$), temperatures ($T$), and viscous shear stress (${\tau _{xy}}$) at the instants of $t = 100\;000\tau ,\;102\;000\tau ,\;104\;000\tau ,...,\;1\;000\;000\tau $. Then, we obtain the final data by averaging all the sample data. In addition, to avoid the influence of boundary conditions, we subsample the data points in the region of $0.1H \le y \le 0.9H$ to form the training dataset.

In this case, the target variable is defined as the viscous shear stress (${\tau _{xy}}$), and the function set is defined as $\left\{ { + , - , \times , \div } \right\}$. The terminal set is almost the same as that in the case of one-dimensional shock wave, i.e., $\left\{ {K{n_\theta },K{n_\rho },\frac{{\partial u}}{{\partial y}},\frac{{\partial \rho }}{{\partial y}},\frac{{\partial \theta }}{{\partial y}},\rho ,\theta ,\mu ,\gamma ,\omega } \right\}$ , except that the high-order gradients are removed. The motivation for removing high-order gradients is that the derived constitutive relation should be applicable in CFD. If containing high-order gradients, the constitutive relation would be unstable and require additional boundary conditions \citep{zhong1993stabilization,bobylev1982chapman,struchtrup2003regularization,torrilhon2004regularized,singh2017derivation}, which are the common problems with the kind of Burnett equations. Besides, note that the gradient-length local (GLL) Knudsen number in this case is defined as $K{n_Q} = \frac{{{\lambda _l}}}{Q}\frac{\partial Q}{\partial y} $.

Similar to the one-dimensional shock wave case, we employ the DHC-GEP, combined with the constraint of the second law of thermodynamics, to discover the underlying constitutive relation based on the data for the cases of $Kn = 0.01,\;0.3$ and $0.4$. The optimal equation is
\begin{equation}
    {\tau _{xy}} =  - 0.247\left( {\frac{5}{{Kn_\rho ^2 + \gamma }} - \frac{{K{n_\theta }}}{{K{n_\rho }}}} \right)\mu \frac{{\partial u}}{{\partial y}},
\label{eq11}
\end{equation}
which is derived with $\alpha=0.7$. The $\mathrm{L}_{S_p}$ for (\ref{eq11}) is 0, which quantitatively indicates that the entropy productions at all data points are non-negative.

\begin{figure}  
\centering  
\includegraphics[width=13.7cm]{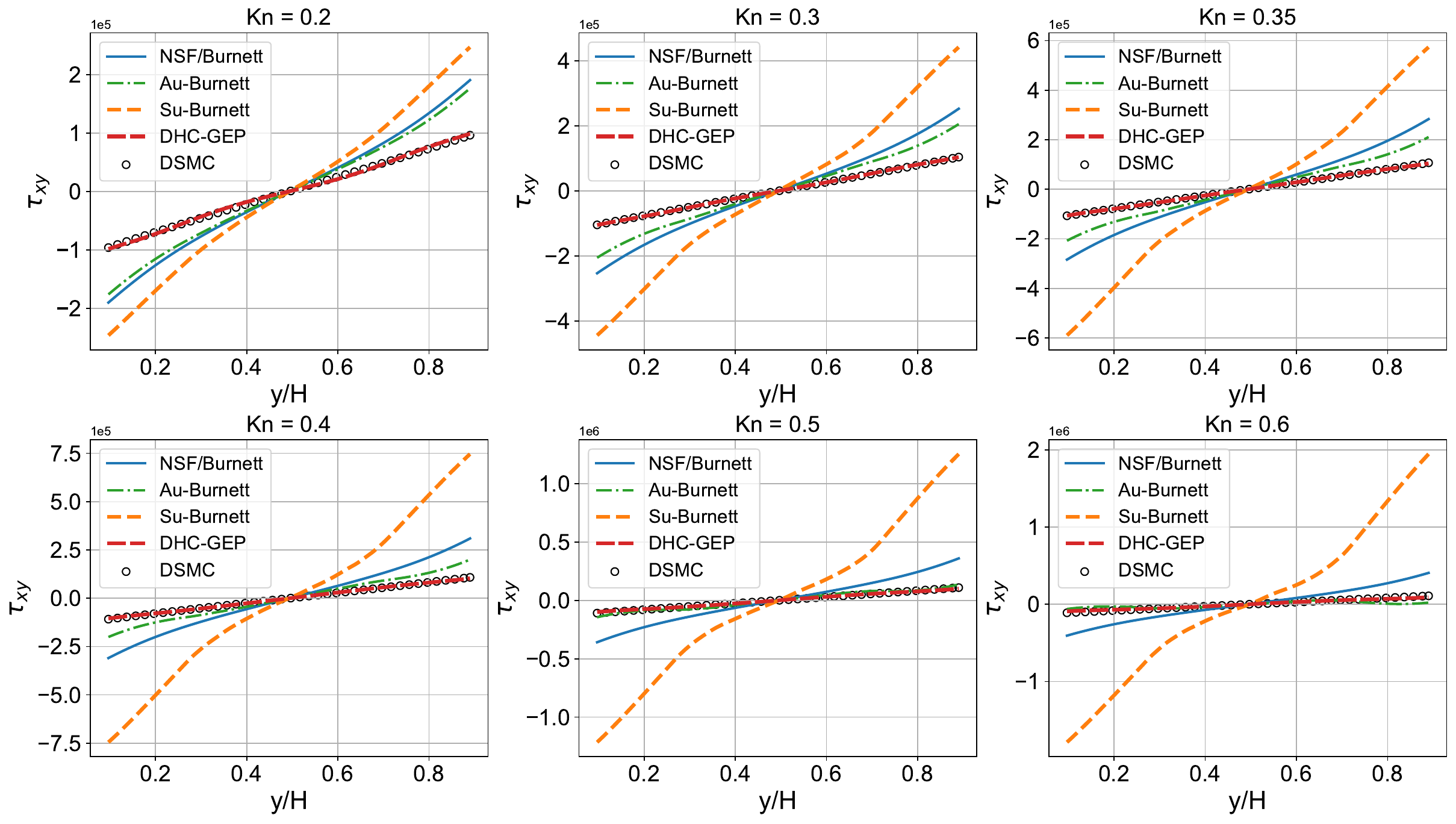}
\caption{Comparisons of the NSF equation, Burnett equation, augmented-Burnett equation (Au-Burnett), super-Burnett equation (Su-Burnett), derived equation of DHC-GEP, and DSMC (exact solution) at $Kn = 0.2,\;0.3,\;0.35,\;0.4,\;0.5$ and $0.6$ for the rarefied Poiseuille flow case. Note that the NSF equation is equal to the Burnett equation in this case.}
\label{fig7}
\end{figure}

We compare the results predicted by (\ref{eq11}) with those predicted by the NSF equation, Burnett equation, augmented-Burnett equation, and super-Burnett equation in figure \ref{fig7}. The relative errors for each constitutive relation are summarized in table \ref{tab4}. It can be found that the derived constitutive relation of DHC-GEP is much more accurate than other equations in a wide range of $Kn$, and is applicable to cases beyond the parameter space of the training data, i.e., $0.4 < Kn \le \;0.6$.

\begin{table}
\centering
\footnotesize
\begin{tabular}{lcccccccc}
& \multicolumn{1}{l}{$Kn= 0.01$} & \multicolumn{1}{l}{$Kn= 0.1$} & \multicolumn{1}{l}{$Kn= 0.2$} & \multicolumn{1}{l}{$Kn= 0.3$} & \multicolumn{1}{l}{$Kn= 0.35$} & \multicolumn{1}{l}{$Kn= 0.4$} & \multicolumn{1}{l}{$Kn= 0.5$} & \multicolumn{1}{l}{$Kn= 0.6$}   \\
DHC-GEP     & 0.028   & 0.083  & 0.044  & 0.019  & 0.025   & 0.039  & 0.076  & 0.137   \\
NSF/Burnett & 0.029   & 0.659  & 0.844  & 1.233  & 1.426   & 1.606  & 1.963  & 2.318   \\
Au-Burnett  & 0.029   & 0.594  & 0.697  & 0.805  & 0.783   & 0.687  & 0.264  & 0.667   \\
Su-Burnett  & 0.030   & 0.921  & 1.435  & 2.948  & 4.012   & 5.306  & 9.045  & 14.000   \\
\end{tabular}
\caption{\label{tab4}Relative errors of the derived equation of DHC-GEP, NSF equation, Burnett equation, augmented-Burnett equation (Au-Burnett), and super-Burnett equation (Su-Burnett) for the rarefied Poiseuille flow case. Note that the NSF equation is equal to the Burnett equation in this case.}
\end{table}

Moreover, note that for continuum flows, the GLL Knudsen numbers ($K{n_\rho }$ and $K{n_\theta }$) approach zero, and (\ref{eq11}) reduces to
\begin{equation}
    {\tau _{xy}} =  - 0.247\left( {\frac{5}{\gamma } - \frac{{K{n_\theta }}}{{K{n_\rho }}}} \right)\mu \frac{{\partial u}}{{\partial y}} =  - 0.247\left( {3 - \frac{{K{n_\theta }}}{{K{n_\rho }}}} \right)\mu \frac{{\partial u}}{{\partial y}}.
\label{eq12}
\end{equation}
Here, we used the value of $\gamma =5/3$ for monatomic gas in the above equation. For the Poiseuille flow in the continuum regime, the pressure along the $y$ direction is observed to remain constant. Taking the flow of $Kn = 0.01$ as an example, the pressure oscillates approximately $0.5\%$ of its absolute value. Considering the ideal gas equation of state $\rho  = \frac{p}{\theta }$, we can obtain
\begin{equation}
    \partial \rho  =  - \frac{p}{{{\theta ^2}}}\partial \theta  =  - \frac{\rho }{\theta }\partial \theta .
\label{eq13}
\end{equation}
Then, combining (\ref{eq13}) with $K{n_\rho } = \frac{{{\lambda _l}}}{\rho }\frac{{\partial \rho }}{{\partial y}}$ and $K{n_\theta } = \frac{{{\lambda _l}}}{\theta }\frac{{\partial \theta }}{{\partial y}}$, we can conclude that $\frac{{K{n_\theta }}}{{K{n_\rho }}} = \frac{{\partial \theta }}{{\partial \rho }} \cdot \frac{\rho }{\theta } =  - 1$. Substituting this relation into (\ref{eq12}) yields
\begin{equation}
    {\tau _{xy}} =  - 0.247\left( {3 + 1} \right)\mu \frac{{\partial u}}{{\partial y}} \approx  - \mu \frac{{\partial u}}{{\partial y}}.
\label{eq14}
\end{equation}
Therefore, it can be concluded that in the continuum regime, the derived constitutive relation of DHC-GEP can be reduced to NSF equations. This makes it applicable to real complex flow problems, where flows tend to be multiscale, i.e., consisting of both continuum flows and non-equilibrium flows.

Finally, we emphasize that as containing only the first-order gradient of velocity, the derived constitutive relation is stable and requires the same boundary conditions as NSF equations. It is convenient to embed it into the well-developed CFD frameworks with minor modification.

\section{Limitations and future works}
\label{sec:limitations}

Although DHC-GEP outperforms Original-GEP in the test cases investigated in this work, it still exhibits certain limitations. Consequently, we objectively outline several limitations associated with DHC-GEP and discuss the potential directions for future optimization as follows.

\begin{itemize}
    \item DHC-GEP remains reliant on the prior knowledge of essential transport coefficients present in the target models. Lacking such knowledge could impede the discovery of meaningful relationships, rendering it impractical to address intricate and unknown problems. Note that dimensional transport coefficients can be derived from fundamental physical property parameters. Therefore, one alternative strategy is incorporating fundamental physical property parameters, such as the molecular mean collision time ($\tau$) and molecular mean free path ($\lambda$), into the terminal set. Our advanced test presented in §\,\ref{sec:diffusion flow}, where the diffusion equation is discovered without incorporating the diffusion coefficient into the terminal set, has preliminarily demonstrated the feasibility of this strategy.
    \item DHC-GEP necessitates computing the derivatives of training data, which can pose a notably ill-conditioned challenge, particularly in the presence of significantly noisy training data. In future investigations, a critical research direction would be to explore the incorporation of the weak formulation utilized in the prior works by \citet{gurevich2019robust}, \citet{reinbold2020using} and \citet{alves2022data} into DHC-GEP. 
    \item As an initial study on DHC-GEP, the two non-equilibrium test cases investigated in this study are one-dimensional, and thus tensor symmetry and rotational invariance have not been considered. The target variables of DHC-GEP in the present work include scalars and one component of vectors (or tensors), which cannot be generalized to discover the equation constituting of tensor variables for higher-dimensional problems. A feasible avenue for future research is to incorporate the dimensional homogeneity constraint into multidimensional gene expression programming (M-GEP) proposed by \citet{weatheritt2016novel}, which modified Original-GEP for the purpose of tensor modelling. Furthermore, in our work, the constraint of dimensional homogeneity is integrated in a non-intrusive manner, which does not alter the fundamental features of Original-GEP, including chromosome structure, expression rules, selection, and genetic operators. Therefore, the constraint of dimensional homogeneity can be conveniently incorporated into other variants of genetic programming such as M-GEP, as long as their individuals can be translated into mathematical expressions.
    \item The denominators of both (\ref{eq7}) and (\ref{eq11}) include the gradient-length local Knudsen number ($K{n_\rho }$ and $K{n_\theta }$). In numerical calculation, these two equations would become unstable and exhibit discontinuous jumps in the regions where density and temperature are uniform and both local Knudsen numbers vanish. One feasible solution is to use a relaxation technique. For instance, (\ref{eq7}) can be relaxed to
    \begin{equation}
        {q_x} =  - 0.444\left( {6K{n_\rho }  + K{n_\theta }+\frac{{K{n_\theta }K{n_\rho }-0.01}}{{Kn_{\rho}^{2}+0.01 }} + \frac{{K{n_\theta }K{n_\rho }-0.01}}{{Kn_{\theta}^{2}+0.01  }}} \right)\kappa \frac{{\partial \theta }}{{\partial x}},
    \label{eq7_relax}
    \end{equation}
    By embedding (\ref{eq7_relax}) into an open-source CFD solver, SU2 \citep{economon2016su2}, we can numerically solve the one-dimensional wave. Figure \ref{numerically_solve_shock} shows the results for a case with freestream Mach number of 3.0. It can be observed that the results based on the DHC-GEP-derived constitutive relation exhibit excellent agreement with DSMC simulations.
    \begin{figure}  
    \centering 
    \includegraphics[width=13cm]{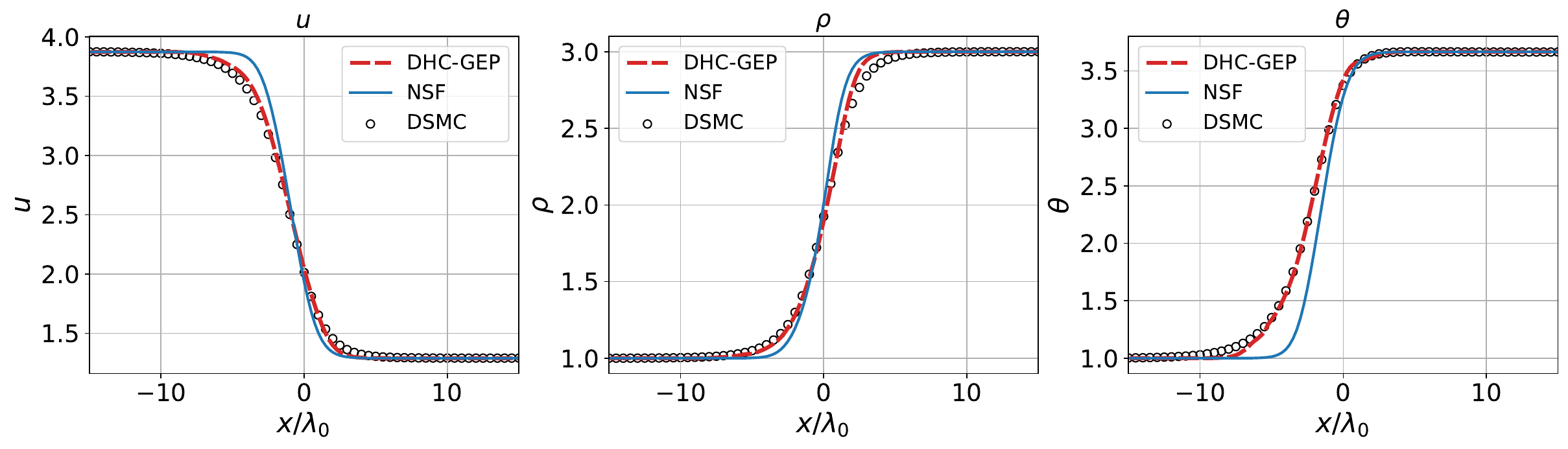}
    \caption{ Comparisons of the results numerically computed with the DHC-GEP-derived/NSF constitutive relations for the one-dimensional shock wave with $M{a_\infty } = 3.0$.} 
    \label{numerically_solve_shock}
    \end{figure}
    It should be noted that this merely serves as a manual correction method and does not fundamentally resolve the issue of possible singularities in the obtained equations. To enable DHC-GEP to automatically yield equations that require no further correction and can be employed in numerical computations, one possible approach is to couple the evolutionary process of DHC-GEP with the numerical computation process. When evaluating the loss function of a particular equation, it involves more than simply computing the discrepancy between the stresses (or heat flux) predicted by the equation and those obtained from DSMC. Instead, the equation would be embedded into the CFD numerical computation program, and the equation is numerically solved. The computed velocity, density, and temperature are then compared with the DSMC results to obtain the discrepancy between them as the loss value for the equation. This strategy provides a possible way to yield equations that require no further corrections and can be utilized for numerical computations, but also imposes higher demands on the computational performance of computers.
    \item It is difficult for the present strategy (singly employing DHC-GEP) to discover a real universal governing equation. For example, the derived constitutive relation in the case of rarefied Poiseuille flow is definitely not universally valid for any non-equilibrium flows. This equation is essentially a model equation for a specific class of flows, the flow characteristics of which are similar to Poiseuille flow, instead of the real universal constitutive relation. Considering the non-equilibrium transport being complex, it is believed that the real universal constitutive relation tends to be correspondingly complex, for instance, containing high-order gradients. Although such constitutive relation is accurate, it is difficult or even impossible to embed it into the present CFD frameworks. Therefore, it is not suitable for practical engineering applications. Alternatively, we should focus on the model equations, each of which is valid for a specific class of problem and is easy to use in practice. One promising direction is combining clustering algorithms \citep{schmid2011applications,callaham2021learning} with DHC-GEP. Based on a complex flow that contains a variety of flow characteristics, clustering algorithms can be first employed to divide the flow into several sub-flows. Then, DHC-GEP is employed in each sub-flow to discover the corresponding model equations. The derived constitutive relations in the present work are two examples of all model equations. During specific numerical computations, for each mesh point, we can first determine which sub-flow it belongs to, and then apply corresponding model equations. It is noteworthy that although the above discussions are based on non-equilibrium constitutive relations, they can also be extended to other fields.
\end{itemize}

\section{Conclusions}
\label{sec:CAD}
In this work, an improved algorithm for gene expression programming is proposed, referred to as dimensional homogeneity constrained gene expression programming (DHC-GEP). The constraint of dimensional homogeneity is introduced to the Original-GEP method through an additional dimensional verification process. The major features of Original-GEP are not changed, including the structure of chromosomes, the rules of expression, selection and reproduction. Therefore, DHC-GEP inherits the advantages of Original-GEP. Specifically, DHC-GEP discovers the forms of functions and their corresponding coefficients simultaneously. The resulting equations are constructed by randomly combining basic elements in the terminal set and function set while satisfying the syntactic requirements of the mathematical expression, rather than by linearly combining the predetermined candidate functions, leading to great flexibility. The chromosomes in DHC-GEP have fixed length, avoiding bloating and unaffordable computational costs that are common in other evolutionary algorithms when dealing with complex problems. On the other hand, the length of open reading frame is variable, ensuring strong expressivity. DHC-GEP is tested on two benchmark cases, including diffusion equation and vorticity transport equation. It is demonstrated that DHC-GEP is capable of discovering the right equations from both the views of data fitting and revealing physical principles, and the result of DHC-GEP is more robust to hyperparameters, the noise level and the size of datasets, compared to that of Original-GEP. When the data are noisy or scarce, Original-GEP tends to converge to the overfitting results with lower loss. However, these overfitting results can be automatically filtered out in DHC-GEP. Moreover, DHC-GEP is more computationally economical than Original-GEP, as DHC-GEP can identify some individuals as invalid individuals through dimensional verification and skip the process of evaluating losses for these individuals. The total cost decreases despite the extra expense of dimensional verification. These advantages make DHC-GEP a promising tool for discovering unknown governing equations from molecular simulation data.

We also present how to employ DHC-GEP to discover the unknown constitutive relations for two representative non-equilibrium flows, including one-dimensional shock wave and rarefied Poiseuille flow. We generate the datasets by DSMC, which does not assume any governing equations. For other scientific and engineering disciplines, the datasets could be generated by experiments or first principle calculations without any assumptions of governing equations. Then, in these investigations for unknown equations, we meticulously design the terminal set to incorporate pertinent physical knowledge, such as Galilean invariance. Besides, the constraint of the second law of thermodynamics is embedded via adding an additional loss term, which is related to entropy production, to the loss function. Finally, based on the terminal and function sets, DHC-GEP conducts a global search in the space of mathematical expressions until a satisfying equation is obtained. For the two cases in our work, the derived constitutive relations are much more accurate than the conventional equations derived based on physics knowledge and phenomenological assumptions (including NSF, Burnett, augmented-Burnett, and super-Burnett equations) in a wide range of Knudsen number and Mach number, and are even applicable to cases beyond the parameter space of the training data. In addition, the physical properties of the derived constitutive relations are excellent. Specifically, the derived constitutive relations contain only the first order gradients, and hence are stable and require the same boundary conditions as NSF equations. As a comparison, the kind of Burnett equations are unstable, and cannot be exactly proven to satisfy the second law of thermodynamics. It is convenient to embed the derived constitutive relations into the well-developed CFD frameworks with minor modifications.

\appendix

\section{Automatic differentiation}\label{AD}

The spatial and temporal derivatives are computed with automatic differentiation (AD), which conducts a non-standard interpretation of a given computer program by replacing the domain of the variables to incorporate derivative values and redefining the semantics of the operators to propagate derivatives per the chain rule of differential calculus \citep{baydin2018automatic}. Theoretically, AD can be implemented using any regression algorithms. In this work, we utilize feedforward neural networks due to their exceptional fitting capabilities, which have been widely validated in various regression problems. In theory, multilayer feedforward neural networks can approximate any complex functional relationship \citep{hornik1989multilayer}.

\begin{figure}  
\centering 
\includegraphics[width=7cm]{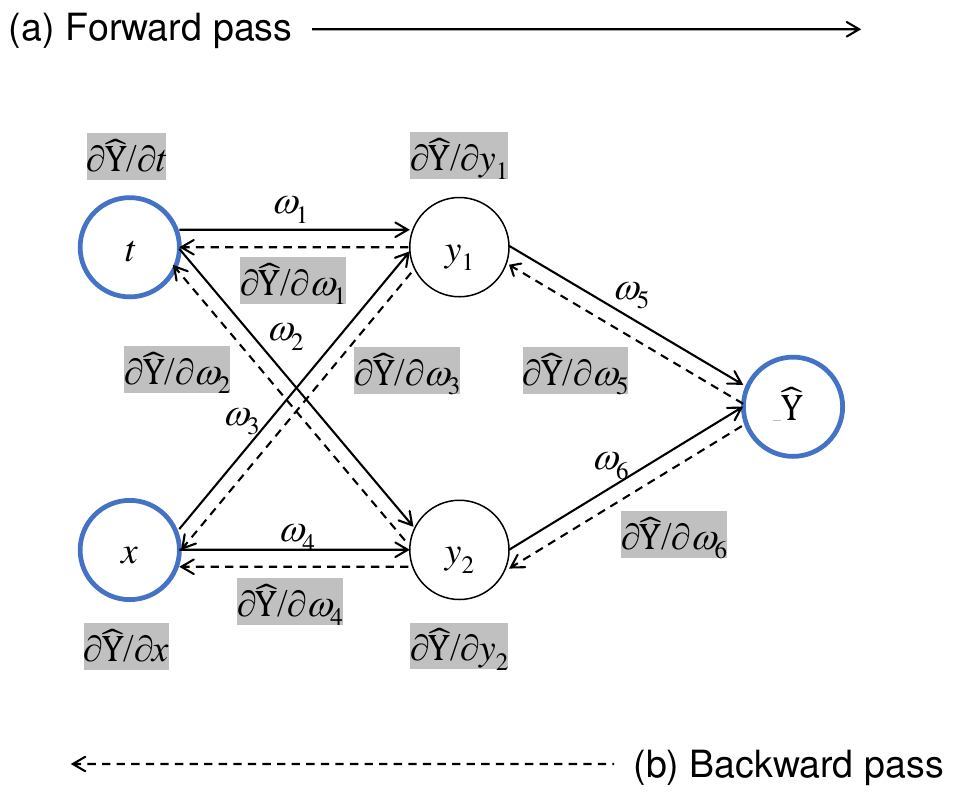}
\caption{Schematic diagram of backpropagation. (a) The input variables are passed forward, and the corresponding target variables are predicted at the output layer. (b) The gradients are propagated backwards using the chain rule of differential calculus.} 
\label{BP}
\end{figure}

Specifically, we begin by constructing a standard feedforward neural network, with spatial coordinates and time as inputs, and the physical variables to be fitted as outputs (such as horizontal velocity ($u$), vertical velocity ($v$), and vorticity ($\omega_z $) in the case of cylinder flow). All the available data are fed into the network for training, and the loss function is defined using the mean squared error (MSE) as
\begin{equation}
    {\mathrm{L_{MSE}}} = \frac{1}{N}\sum\limits_{i = 1}^N { \left ( {{\hat Y}_i} - {Y_i} \right )^2  } ,
\label{app_MSE}
\end{equation}
where the variable with a superscript $\wedge$ is the predicted variable, and $N$ is the total number of data points. Once training is completed, the derivatives of the physical variables with respect to the input variables can be obtained by the backwards propagation of the sensitivity of the objective value at the output layer, utilizing the chain rule of differential calculus (see in figure \ref{BP}).

AD calculates the derivatives by accumulating values during code execution to produce numerical derivative evaluations, instead of derivative expressions. This method enables precise evaluation of derivatives at machine precision, with only a small constant factor of overhead and ideal asymptotic efficiency \citep{baydin2018automatic}. It is reported in \citet{raissi2018deep} and \citet{xu2019dl} that AD is more stable than finite differences and polynomial interpolation with respect to noise, which is beneficial for discovering the correct equations from noisy data. Another interesting point is that based on the trained neural networks, it is also convenient to combine the output variables and compute the derivatives of their combination with respect to input variables. Therefore, while not validated in this work, it is promising to incorporate partial differential operators into the function set by coupling the trained neural networks and DHC-GEP.

In this work, the feedforward neural networks have 8 hidden layers, each with 20 neurons. During the training process, Adam optimization \citep{kingma2014adam} is employed for the first 20 000 iterations, after which L-BFGS-B optimization \citep{byrd1995limited} is utilized until convergence.

\begin{figure}  
\centering 
\includegraphics[width=12cm]{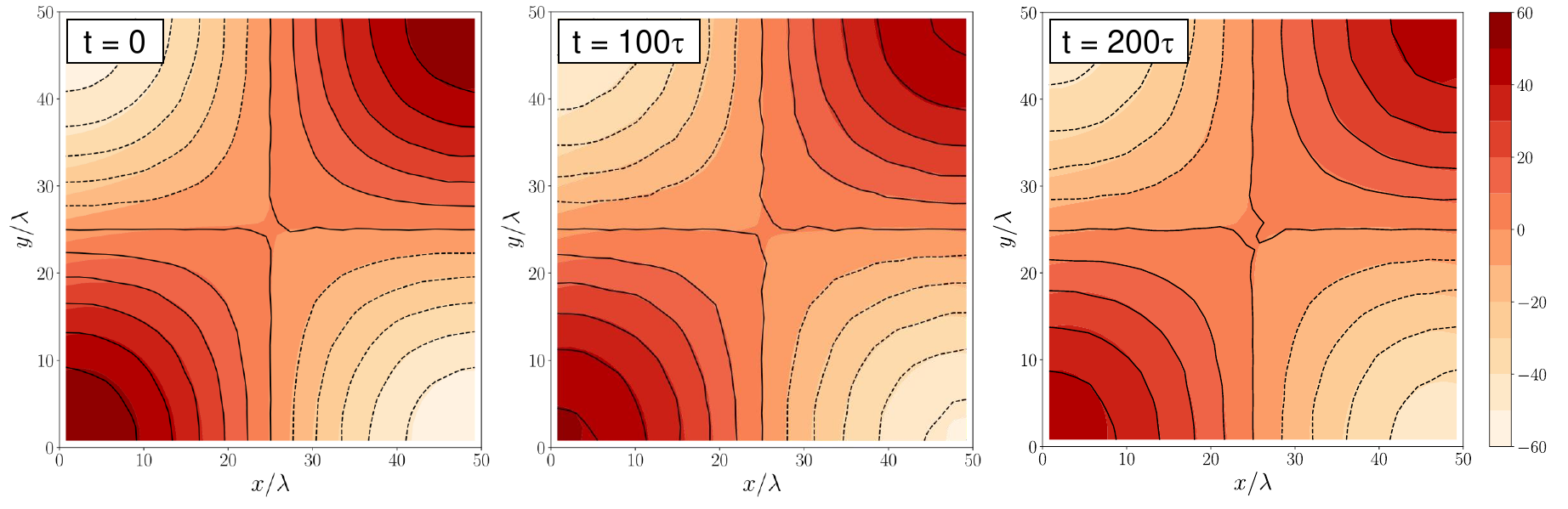}
\caption{The theoretical values of $\frac{\partial^2 \omega_z }{\partial x^2}$ and the numerical values computed with automatic differentiation at the instants of $t = 0, 100\tau, 200\tau$. The contour lines represent the theoretical values, while the colour regions represent the numerical values obtained through automatic differentiation.} 
\label{Tp}
\end{figure}

Moreover, to demonstrate the accuracy of the derivatives computed with this method, we conduct a quantitative analysis based on the DSMC data of Taylor-Green vortex. Specifically, according to the analytical solution of the Taylor-Green vortex given in (\ref{eq4}), we can theoretically obtain the higher-order derivative term $\frac{\partial^2 \omega_z }{\partial x^2}$ as follows:
\begin{equation}
    \frac{\partial^2 \omega_z }{\partial x^2} =2v_{0}\cdot {\rm cos}\left (x \right ){\rm cos}\left (y \right ) {\exp \left( { - 2\upsilon t} \right)} = -\omega_z.
\label{eq_Taylor_proof}
\end{equation}
Figure \ref{Tp} shows the theoretical values of $\frac{\partial^2 \omega_z }{\partial x^2}$, which equals $-\omega _z$, and the numerical values computed with automatic differentiation at the instants of $t = 0, 100\tau, 200\tau$.

At these three instants, the relative errors between the numerical and theoretical values are 0.04, 0.03, and 0.03, respectively. Hence, this analysis preliminarily indicates that the precision of the high-order derivatives obtained through automatic differentiation is satisfactory.

In addition, we generate the accurate flow field of Taylor-Green vortex based on (\ref{eq4}), and then add noise of varying magnitudes into the accurate data. Specifically, the definition of noise data is consistent with the approach in \citet{gurevich2021learning}. Taking vorticity as an example, the vorticity data with noise is given by
\begin{equation}
    \omega_{z,\sigma } = \omega_{z} + \sigma N_1 s_{\omega}.
\label{eq_noise}
\end{equation}
Here, $\sigma$ is the noise level, $N_1$ is the sampled standard normal random variables at each point in space and time, and $s_{\omega}$ is the sample standard deviation of $\omega_{z}$. Based on the noisy data, we use AD to compute the derivatives and use DHC-GEP to discover the underlying governing equations. Table \ref{tab_noise} displays the relative error of AD-computed $\frac{\partial^2 \omega_z }{\partial x^2}$ compared to theoretical values, along with the corresponding derived governing equations. It can be observed that when the noise level does not exceed $8\%$, AD is capable of accurately computing second-order derivative terms, and DHC-GEP can discover correct governing equations. \citet{xu2019dl} conducted a similar study, and they found that with the assistance of the AD based on neural network without imposing additional constraints, the sparse regression could discover the correct equations from noisy data with a noise level of around $10\%$. In comparison, the sparse regression using polynomial interpolation, as described by \citet{rudy2017data}, could, at best, discover the correct equations from data with a noise level of around $1\%$.

\begin{table}
\centering
\begin{tabular}{ccc}
Noise level & Relative error ($\frac{\partial^2 \omega_z }{\partial x^2}$) & Derived equation  \\
2\%         & 0.053            & $\frac{{\partial {\omega _z}}}{{\partial t}} = 1.048\upsilon \left( {\frac{{{\partial ^2}{\omega _z}}}{{\partial {x^2}}} + \frac{{{\partial ^2}{\omega _z}}}{{\partial {y^2}}}} \right)$                 \\
4\%         & 0.054            & $\frac{{\partial {\omega _z}}}{{\partial t}} = 1.054\upsilon \left( {\frac{{{\partial ^2}{\omega _z}}}{{\partial {x^2}}} + \frac{{{\partial ^2}{\omega _z}}}{{\partial {y^2}}}} \right)$                 \\
6\%         & 0.058            & $\frac{{\partial {\omega _z}}}{{\partial t}} = 1.040\upsilon \left( {\frac{{{\partial ^2}{\omega _z}}}{{\partial {x^2}}} + \frac{{{\partial ^2}{\omega _z}}}{{\partial {y^2}}}} \right)$                 \\
8\%         & 0.063            & $\frac{{\partial {\omega _z}}}{{\partial t}} = 1.033\upsilon \left( {\frac{{{\partial ^2}{\omega _z}}}{{\partial {x^2}}} + \frac{{{\partial ^2}{\omega _z}}}{{\partial {y^2}}}} \right)$                 \\
10\%        & 0.134            & $\frac{{\partial {\omega _z}}}{{\partial t}} = 1.006\upsilon \left( {-\frac{{{\partial ^2}{\omega _z}}}{{\partial {x^2}}} + 3\frac{{{\partial ^2}{\omega _z}}}{{\partial {y^2}}}} \right)$                
\end{tabular}
\caption{\label{tab_noise}Relative error of AD-computed $\frac{\partial^2 \omega_z }{\partial x^2}$ and the corresponding derived governing equations at different noise levels.}
\end{table}

In general, given the remarkable fitting capabilities of neural networks, it is not difficult to achieve accurate fitting and derivative calculations, as long as the training data can represent the gradients. Nonetheless, it is imperative to clarify that if the training data are overly scarce and noisy, the learned neural networks would fail to calculate the derivatives correctly.

\section{Dc domain}\label{Dc domain}

\begin{figure}  
\centering  
\includegraphics[width=12cm]{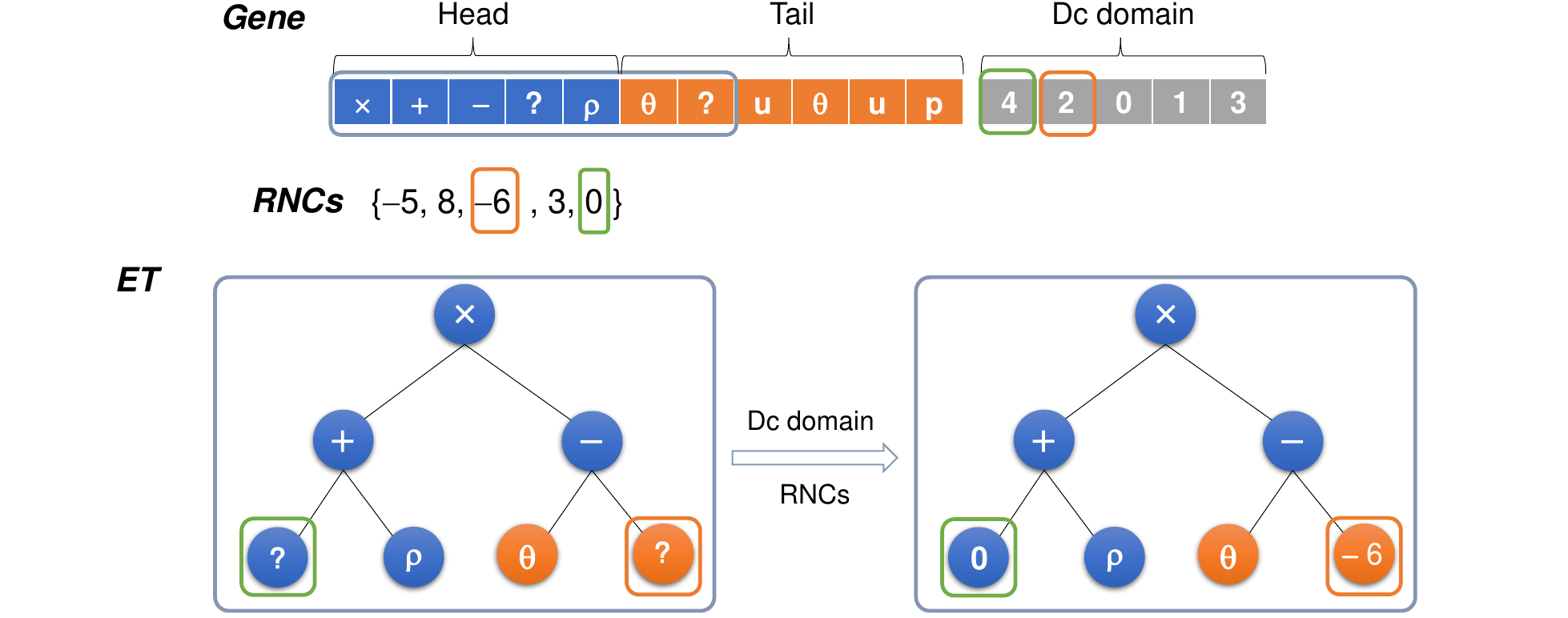}
\caption{Schematic diagram of how Dc domain works.} 
\label{su_fig4}
\end{figure}

It is common to have numerical constants in governing equations. In the gene of GEP, the numerical constants are represented by a special terminal "?". During expression, the "?" would be replaced with the random numerical constants (RNCs) from a predefined array, under the guidance of a special domain called Dc domain. Dc domain is an additional domain behind the gene, and consists of the indices that determine which RNCs are selected to replace the "?". In this work, RNCs are randomly integers between -10 and 10. Indeed, RNCs could be selected from a broader range and are not inherently limited to integers. Nevertheless, considering the utilization of the linear scaling technique, such an extension appears unnecessary. This is because the multiplication of the scaling factor and RNCs allows for the generation of arbitrary numerical constants, which is introduced in Appendix \ref{linear scaling}. Consequently, it is a common practice within the GEP community to define RNCs as integers within the specified range of (-10, 10). A simple schematic diagram is shown in figure \ref{su_fig4}. It is worth noting that, due to the introduction of negative constants, the subtraction operator in the function set is redundant. The inclusion of the subtraction operator in this work is simply to maintain consistency with other researches relevant to GEP.

In GEP, there are several genetic operators specifically for Dc domain and RNCs, including Dc inversion, Dc transposition, and random constant mutation operators. These genetic operators have similar mechanisms to the genetic operators introduced in §\,\ref{sec:General flowchart}, and are also invoked with certain probabilities, as listed in table \ref{Possibilities_Dc}.

\begin{table}
\centering
\begin{tabular}{lc} 
Genetic operator          & Probabilities  \\ 
Dc specific mutation      & 0.05           \\
Dc specific inversion     & 0.1            \\
Dc specific transposition & 0.1            \\
Random constant mutation  & 0.02           \\
\end{tabular}
\caption{\label{Possibilities_Dc}Probabilities for genetic operators of Dc domain being invoked.}
\end{table}

\section{Linear scaling}\label{linear scaling}

Generally, GEP discovers the structure of derived equations quickly, but struggles in optimizing the numerical constants. For example, if the target equation is $y=9.7x^{2}$ (where $y$ is the target variable and $9.7x^{2}$ is the mathematical expression that we want to derive), GEP would discover $x^{2}$ easily but struggles in discovering the numerical constant (9.7), because the number space is relatively large. To address this issue, a linear scaling technique was proposed in \citep{keijzer2003improving}. Specifically, assuming that the mathematical expression of a chromosome is $g\left( x \right)$, then the scaling factor ($\xi$) can be directly obtained as
\begin{equation}
    \xi =\frac{y}{g\left ( x \right ) }.
\label{linear scaling equation}
\end{equation}
Revisiting the previous example, as long as GEP generates a chromosome with its mathematical expression being $ax^{2}$, where $a$ can be any integer constant, the correct equation can be discovered simultaneously since the scaling factor ($9.7/a$) of this chromosome can be easily obtained using (\ref{linear scaling equation}). In this fashion, GEP focuses on discovering the function structures, which is its forte, and the linear scaling technique facilitates the identification of the appropriate scaling factors.

It is important to note that in the example provided above, for the sake of simplification, $g\left( x \right)$ is represented as a simple monomial. However, it can be a linear combination of multiple functional terms, such as $g\left( x \right) = ax^3+bx^2+cx$. Here, $a$, $b$, and $c$ are integer constants, which are generated using the rules introduced in Appendix \ref{Dc domain}. The final equation is $y=\xi g\left( x \right) = a\xi x^3+b\xi x^2+c\xi x$. Hence, the numerical constants are determined by a joint influence of the scaling factor ($\xi$) and the integer constants within the chromosome.

\section{Loss function}\label{app_loss_function}

In this work, to evaluate the performances of individuals, we define the loss function using the mean relative error (MRE) as
\begin{equation}
    {\mathrm{L_{MRE}}} = \frac{1}{N}\sum\limits_{i = 1}^N {\left| {\frac{{{{\hat Y}_i} - {Y_i}}}{{{Y_i}}}} \right|} .
\label{app_loss_0}
\end{equation}
Here, the variable with a superscript $\wedge$ is the predicted variable, and $N$ is the total number of data points. In MRE, each data point receives equal attention, making it highly suitable for this study. Since the constitutive relations reflect local momentum and energy transport characteristics, it is imperative to treat each data point with equal attention. 

Moreover, we emphasize that the selection of loss function should be jointly determined based on the distribution of training data and the requirements of problems. It is challenging to ascertain that a particular loss function is always the optimal choice for any given problem. MRE is suitable for this study, but it is not widely applicable to other problems. MRE is sensitive to outliers in the data. A single outlier can significantly increase the relative error, affecting the stability and performance of the model during training. In problems where strict equal attention to each data point is not needed, the Huber loss \citep{huber1992robust} is a better alternative, which is better equipped to handle outliers. 

\section{Hyperparameters}\label{appB}

The three key hyperparameters in the GEP method are the length of head, the number of genes in a chromosome, and the number of individuals in a population. The length of head and the number of genes in a chromosome determine the upper limit of the complexity of the derived equation.  The number of individuals in a population determines the diversity of individuals in a population. Generally, a larger population means a greater diversity of individuals, and a higher computational cost in an evolution as well. According to the parametric study on the diffusion flow case (§\,\ref{sec:diffusion flow}), we find that Original-GEP is quite sensitive to these three hyperparameters. However, this is not the issue of DHC-GEP. Therefore, we keep all hyperparameters consistent across all cases. Specifically, the length of head is set to 15, and the number of genes a chromosome is set to 2, ensuring that the upper limit of the complexity of equations DHC-GEP could explore is sufficiently high. The number of individuals in a population is set to 1660 to ensure that there are enough varieties of equations in each generation.

The possibilities of the genetic operators being selected refer to \citet{ferreira2006gene}, listed in tables \ref{su_tab4} and \ref{Possibilities_Dc}, which were concluded from various examples via the trial and error approach. 

\section{Proof of Galilean invariance}\label{appC}

Galilean invariance is a fundamental property of physical laws, which has been proven to be important for constructing constitutive relations with data-driven methods \citep{li2021learning,huang2021learning,han2019uniformly,han2020alternative}. Specifically, this means that the equation forms of physical laws remain invariant in all inertial frames. Assuming that one inertial frame ($x'y'z'$) moves at a constant speed (${\mathbf{v}_0}$) with respect to another inertial frame ($xyz$), the transformations of spatiotemporal coordinates between these two frames are 
\begin{equation}
    \begin{dcases} {\begin{array}{*{20}{l}}
        {\mathbf{r'} = \mathbf{r} - {\mathbf{v}_0}t} \vspace{0.8ex}\\
        {t' = t}
        \end{array}} \end{dcases}.
\label{eq18}
\end{equation}
Here, $\mathbf{r}$ and $\mathbf{r'}$ are the radius vectors. Moreover, the macroscopic state variables satisfy
\begin{equation}
    \begin{dcases} {\begin{array}{*{20}{l}}
        {\mathbf{v'} = \mathbf{v} - {\mathbf{v}_0}} \vspace{0.8ex}\\
        {\rho ',\;\theta ',\;p' = \rho ,\;\theta ,\;p}
        \end{array}} \end{dcases}.
\label{eq19}
\end{equation}
The partial differential operators have the following relations:
\begin{equation}
    \begin{dcases}\begin{array}{*{35}{l}}
        {\frac{{\mathbf{\partial \mathbf{U}}}}{{\partial t'}} = \frac{{\mathbf{\partial \mathbf{U}}}}{{\partial t}} \cdot \frac{{\partial t}}{{\partial t'}} + \frac{{\mathbf{\partial \mathbf{U}}}}{{\mathbf{\partial r}}} \cdot \frac{{\mathbf{\partial r}}}{{\partial t}} \cdot \frac{{\partial t}}{{\partial t'}} = \frac{{\mathbf{\partial \mathbf{U}}}}{{\partial t}} + {\mathbf{v}_0}\frac{{\mathbf{\partial \mathbf{U}}}}{{\mathbf{\partial r}}}} \vspace{1.5ex}\\
        {\frac{{\mathbf{\partial \mathbf{U}}}}{{\mathbf{\partial r'}}} = \frac{{\mathbf{\partial \mathbf{U}}}}{{\partial t}} \cdot \frac{{\partial t}}{{\mathbf{\partial r'}}} + \frac{{\mathbf{\partial \mathbf{U}}}}{{\mathbf{\partial r}}} \cdot \frac{{\mathbf{\partial r}}}{{\mathbf{\partial r'}}} = \frac{{\mathbf{\partial \mathbf{U}}}}{{\mathbf{\partial r}}}} \vspace{1.5ex}\\
        {\frac{{{\partial ^n}\mathbf{U}}}{{\partial {{\mathbf{r'}}^n}}} = \frac{\partial }{{\mathbf{\partial r'}}}\left( {\frac{{{\partial ^{n - 1}}\mathbf{U}}}{{\partial {{\mathbf{r'}}^{n - 1}}}}} \right) = \frac{\partial }{{\mathbf{\partial r}}}\left( {\frac{{{\partial ^{n - 1}}\mathbf{U}}}{{\partial {{\mathbf{r'}}^{n - 1}}}}} \right) =  \cdots  = \frac{{{\partial ^n}\mathbf{U}}}{{\partial {\mathbf{r}^n}}}}
    \end{array}\end{dcases},
\label{eq20}
\end{equation}
where $\mathbf{U}$ represents variables that are relevant to $\mathbf{r}$ and $t$. 

Assuming that the constitutive relation for viscous stress contains velocity ($\mathbf{v}$) explicitly outside the partial differential operators, a simple but representative example is 
\begin{equation}
    \mathbf{\tau}  = \mathbf{v}\frac{{\partial \mathbf{U}}}{{\partial \mathbf{r}}}.
\label{eq21}
\end{equation}
In the inertial frame ($x'y'z'$), it can be derived that
\begin{equation}
    \mathbf{\tau '} = \mathbf{v'}\frac{{\partial \mathbf{U}}}{{\partial \mathbf{r'}}} = \left( {\mathbf{v} - {\mathbf{v}_0}} \right)\frac{{\partial \mathbf{U}}}{{\partial \mathbf{r}}} \ne \mathbf{\tau} ,
\label{eq22}
\end{equation}
which means that such a constitutive relation does not satisfy the Galilean invariance.

On the contrary, if the constitutive relations do not contain velocity ($\mathbf{v}$) explicitly outside the partial differential operators, it is straightforward to prove that they would remain invariant in different inertial frames, i.e., satisfying the Galilean invariance, according to (\ref{eq19}-\ref{eq20}).

\backsection[Acknowledgements]{The authors thank D.A. Lockerby for providing stimulating discussions.}

\backsection[Funding]{This work was supported by the National Natural Science Foundation of China (Grant Nos. 92052104 and 12272028). The results were obtained on the Zhejiang Super Cloud Computing Center M6 Partition.}

\backsection[Declaration of interests]{The authors report no conflict of interest.}

\backsection[Data availability statement]{The datasets and source codes used in this work are available on GitHub at \href{https://github.com/Wenjun-Ma/DHC-GEP}{https://github.com/Wenjun-Ma/DHC-GEP}.}

\backsection[Author contributions]{W.M., J.Z. and D.W. contributed to the ideation and design of the research; W.M., K.F. and H.X. generated the datasets; W.M. performed the research; W.M., J.Z. and D.W. wrote the paper.}

\bibliographystyle{jfm}


\end{document}